# Towards high resolution, validated and open global wind power assessments

E. U. Peña-Sánchez[1,2,†], P. Dunkel[1,2,†,*], C. Winkler[1,2,†], H. Heinrichs[1], F. Prinz[1], J.M. Weinand[1], R. Maier[1,2], S. Dickler[1], S. Chen[1,3], K. Gruber[4], T. Klütz[1], J. Linßen[1], and D. Stolten[1,2]

1 Forschungszentrum Jülich GmbH, Institute of Climate and Energy Systems, Jülich Systems Analysis, 52425 Jülich, Germany

2 RWTH Aachen University, Chair for Fuel Cells, Faculty of Mechanical Engineering, 52062 Aachen, Germany

3 Forschungszentrum Jülich GmbH, Institute of Bio- and Geosciences - Agrosphere (IBG-3), 52425 Jülich, Germany

4 Institute for Sustainable Economic Development, University of Natural Resources and Life Sciences, Vienna, Austria

[†] Equally contributed
*corresponding author: p.dunkel@fz-juelich.de

## Abstract

Wind power is expected to play a crucial role in future net-zero energy systems, but wind power simulations to support deployment strategies vary drastically in their results, hindering reliable design decisions. Therefore, we present a transparent, open source, validated and evaluated, global wind power simulation tool called *ETHOS.RESKit$_{Wind}$* with high spatial resolution and customizable designs for both onshore and offshore wind turbines. The tool provides a comprehensive validation and calibration procedure using over 16 million global measurements from metrerological masts and wind turbine sites. We achieve a global average capacity factor mean error of 0.006 and Pearson correlation of 0.865. In addition, we evaluate its performance against several aggregated and statistical sources of wind power generation. The release of *ETHOS.RESKit$_{Wind}$* is a step towards a fully open source and open data approach to accurate wind power modeling by incorporating the most comprehensive simulation advances in one model.

# Introduction

Wind power is placed as one of the largest renewable sources for the upcoming decades [1–4]. Thus, evaluating wind power resources is essential to develop strategies for the energy systems transformation, for instance in capacity planning, designing adequate market frameworks, or for increasing the speed of planning and permitting [2,4–6]. Being able to accurately assess wind resources ultimately leads to more reliable future energy transformation strategies.

Wind power resources depend on the location (spatial dependency), on the conditions at a particular time (temporal dependency), and on the wind turbine performance (technology dependency) to translate wind speed's kinetic energy into electricity output. Incorporating these three aspects in one wind energy assessment tool is essential to enhance the robustness and reliability of results. In addition, the validation of results is necessary to evaluate the performance of the model. There have been continuous efforts within the renewable energy simulation community to capture these dependencies using time-resolved and geospatially-constrained wind power simulation models [7].

Widely used, state-of-the-art, open-source models that account for the three wind power dependencies are *Renewables.ninja* [8], *RESKit* [9], *pyGRETA* [10] and *Atlite* [11] (see Table 1). The first three models use weather data based on MERRA-2 [12], and *RESKit* and *pyGRETA* take advantage of the higher-resolved Global Wind Atlas (GWA) [13] to increase the spatial resolution to 1 $km^2$. *Atlite* employs ERA5 [14] data, which compared to MERRA-2 has a higher spatial resolution of 0.28° [~ 31 $km^2$] and offers wind speeds at 100 m height instead of 50 m as in the case of MERRA-2.

**Table 1: Comparison of common global open-source wind energy models**

| Model | Author(s), year | Data source | Resolution [time, lat/lon] | Turbine modeling characteristics | Validation of results |
|---|---|---|---|---|---|
| *Renewables.ninja* | Staffell & Pfenninger[8], 2016 | MERRA-2 | 1 hour, 0.5°/0.625° | 141 existing turbine models | Time-resolved country-level aggregated data in eight European countries |
| *RESKit* | Ryberg et al.[9], 2017 | MERRA-2 | 1 hour, 0.5°/0.625° (scaled to 1 $km^2$) | 123 existing turbine models and user-defined configurations on hub height, rotor diameter and capacity to derive synthetic power curves | Two hourly-resolved wind park generation data: one in France and one in the Netherlands and monthly power generation in Denmark |
| *pyGRETA* | Siala & Houmy[10], 2020 | MERRA-2 | 1 hour, 0.5°/0.625° (scaled to 1 $km^2$) | Allows user-defined changes in cut-in and cut-out wind speeds and the full-load stage in the power curves | Not provided |
| *Atlite* | Hofmann et al.[11], 2021 | ERA5 | 1 hour, 0.28125° | 27 existing turbine models | Not provided |

All four models face two primary limitations: First, the absence or restricted availability of validation procedures and, second, the unaddressed inherited mean errors from their weather data source as shown by various studies [15–22]. The most overlooked aspect is the validation of model outcomes despite its crucial relevance to narrow uncertainties and enhancing the robustness of assessments as emphasized by other authors [7,8]. Only *Renewables.ninja* and *RESKit* provide a validation procedure at all, but exclusively for European wind production. To the best of the authors' knowledge, no supplementary validations of these models have been conducted in other regions. Consequently, an evaluation of the models' reliability and performance on a global scale remains an open question. *Renewables.ninja* validated their model results against monthly-aggregated country wind power generation data from ENTSO-E as well as nationally aggregated wind power generation data with at least hourly resolution from eight power system operators for eight European countries. The authors [8] found a systematic mean error in wind speeds in MERRA-2 across Europe. Based on this finding, they calibrated the results of their model by incorporating "national correction factors". The *RESKit* model [9] has been validated against hourly power generation data from two wind parks, resulting in a high Pearson correlation between 0.80-0.88, with total power generation underestimations between 5 and 37%. In addition, the model performance was compared with monthly power generation data from 86 turbines in Denmark, where the majority of deviations in power generation range from -20% to 30%.

The second limitation is defined by the absence or insufficient measures taken to rectify mean errors present in the input data. Previous studies have documented that reanalysis data inherently contain certain deviations and mean errors. For instance, seasonal and diurnal mean errors in MERRA-2 and ERA5 wind speed data have been found previously [21], as well as terrain-related deviations when comparing wind speeds from reanalysis data with wind speed measurements [20,18] .Furthermore, statistical comparisons of these two datasets have been conducted [16] [19], ultimately concluding that ERA5 exhibited superior performance in comparison to MERRA-2. In addition, global mean errors in wind speeds at 10m were also found in the GWA (see Supplementary material 1.12) in particular an overestimation in wind speeds in intertropical regions, Mediterranean Europe, the western half of the USA and the southern hemisphere, and an underestimation in the northern hemisphere of the Eurasian plate and the eastern half of the USA and Canada (see Supplementary material Figure 1). Therefore, the evaluation and subsequent correction of wind speeds derived from reanalysis data can contribute significantly to the accuracy of wind energy assessments. Notably, although *Renewable.ninja* and *RESKit* acknowledge such effects and indirectly address them via their validation procedure, none of the listed models has utilized wind speed correction measures to address inherent mean errors in reanalysis data.

To cover the existing bandwidth of wind turbine characteristics, simulating as many commercially available wind turbines as possible can support achieving more realistic power generation estimations. As presented in Table 1, most models offer to simulate the performance of such turbines although the available number varies from 27 to 141. The most flexible approach when it comes to user-defined turbines is provided by *RESKit* [9] because it is the only model that allows the user to define a synthetic wind power curve, in addition to the ones declared by the manufacturers, based on three wind turbine parameters: hub height, rotor diameter and capacity. This is especially useful for simulating prospective wind turbines, which is often necessary when evaluating future scenarios. In summary, the identified constraints in the reviewed literature comprise the lack of thorough validation encompassing regions beyond Europe, the absence of mean error corrections in the input weather data source, and the lack of incorporating and evaluating the performance of contemporary and prospective wind turbine models.

In this article, we address the above-mentioned limitations of wind power models to enhance their reliability and applicability. Thus, this study introduces *ETHOS.RESKit$_{Wind}$*, a novel wind power model based on *RESKit*. Our model addresses the identified limitations through extensive validation, global applicability, and the incorporation of more than 800 wind turbine models. To enhance precision, we implement a comprehensive calibration of wind speed data gathered from 213 global weather mast locations in 25 different countries globally, spanning over 8 million hours of observation after filtration. Furthermore, we validate the simulated wind power output by comparing it with the actual hourly output from 152 turbines and wind farm sites. Finally, we further validate our model by comparing the outcomes with publicly available country-level hourly wind power generation data, as well as with yearly wind power generation estimates derived from statistical analysis. In response to this analysis, we introduce a methodology and provide global correction factors as open data to enhance alignment with widely available country-specific wind power generation data. Through these rigorous measures, our work significantly contributes to the reliability of future wind power simulations. This contribution is of utmost relevance for the ongoing energy transformation, providing a robust foundation for accurate, open and globally applicable wind energy assessments.

# Results

## Wind speed calibration impact and improvements

The calibration of input wind speeds yielded enhanced performance across all wind dependencies. Figure 1 illustrates the impact of wind speed calibration, showing changes when applying the value-based wind speed adjustment. Wind speeds below 3.2 m/s are adjusted upwards, while wind speeds above 3.2 m/s are adjusted downwards by around 13% on average with the largest relative correction at 13 m/s. The calibration effectively reverses an observed overestimation of wind speed in the range relevant for wind turbine power generation (3-25 m/s). This dependency shows a differential correction across wind speed values, highlighting the adequacy of a value-based wind-speed calibration approach. It is important to note that there is a steep reduction in the number of available observations at wind speeds ~20 m/s or higher, which contributes to the fluctuations seen in Figure 1. Additionally, the measured wind speeds exhibit a skewed normal distribution centered around 6 m/s, which is a relatively low average wind speed for wind energy installations.

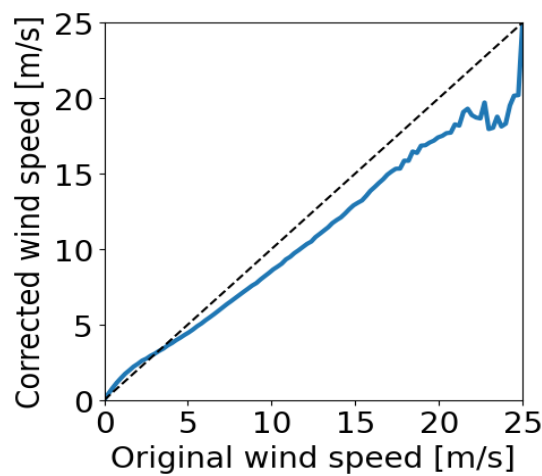

**Figure 1: Effect of the value-based wind speed calibration for different wind speeds**

**Temporal dimension.** The calibration offsets over-representation of high-capacity factors in uncalibrated workflows by addressing wind speed corrections, reducing statistical errors, and ensuring closer alignment with measurements (see Table 2 and Figure 2). The calibration procedure reduces the capacity factor mean error by 11.2% and improves temporal correlation metrics such as root mean square error, Pearson correlation, detrended cross-correlation coefficient, and Perkins' skill score (see Table 2). Despite larger deviations for higher capacity factors, the calibrated workflow *ETHOS.RESKit$_{Wind}$* achieves near-parity in total cumulative electricity generation.

Table 2: Comparison of capacity factors key statistical indicators from two workflow configurations: calibrated and non-calibrated

| Indicator [unitless] | Calibrated | Non-calibrated | Delta (absolute) [%] | Significance for wind energy assessments |
|---|---|---|---|---|
| **Measured mean** | 0.363 | | - | - |
| **Mean** | 0.368 | 0.48 | -11.2 | Closer approximation to the total power generation by the turbines allowing for more precise economic estimations such as levelized cost of electricity, return of investment, value of loss load, etc. |
| **Mean error** | 0.006 | 0.118 | | |
| **Perkins skill score** | 0.87 | 0.82 | +5.0 | Closer approximation to the power generation stochastic variability allowing for more precise technical considerations design to reduce this type of variability in energy systems such as infrastructure capabilities in storage, transmission, etc. |
| **Root-mean square error** | 0.175 | 0.229 | -5.4 | Closer approximation to the power generation natural variability allowing for more precise technical design considerations to optimize power dispatch in energy systems such as infrastructure capabilities in power generation, demand control, system synergies, sector coupling etc. |
| **Pearson correlation** | 0.865 | 0.85 | +1.5 | |
| **Detrended cross-correlation analysis (DCCA) coefficient** | 0.819 | 0.798 | +2.1 | |
| **Count [Million h]** | 7.7 | | - | - |

Zero capacity factors occur at a similar rate (~6-7%) in both measured and simulated workflows, as wind speeds frequently fall below or exceed turbine operational thresholds (see Figure 2). The calibration procedure has a negligible effect on these occurrences. The calibrated *ETHOS.RESKit$_{Wind}$* aligns closely with measured values in the (0-3] capacity factor bin, the most frequent category. In contrast, the uncalibrated workflow underestimates occurrences by about one-third, indicating weaker temporal correlation and probability density alignment. In mid-range capacity factor bins (3-48%), the calibrated workflow aligns better with measured trends despite initially overestimating values and then declining more sharply. Conversely, in high-capacity factor bins (51-99%), the uncalibrated workflow tracks measurements more closely, though these bins contribute less to total electricity generation due to lower cumulative occurrences. At full turbine power (100% capacity factor), both workflows significantly overestimate measurements. However, the uncalibrated workflow overshoots by 2.8 times compared to 0.4 times for the calibrated workflow, significantly impacting total generation and statistical indicators.

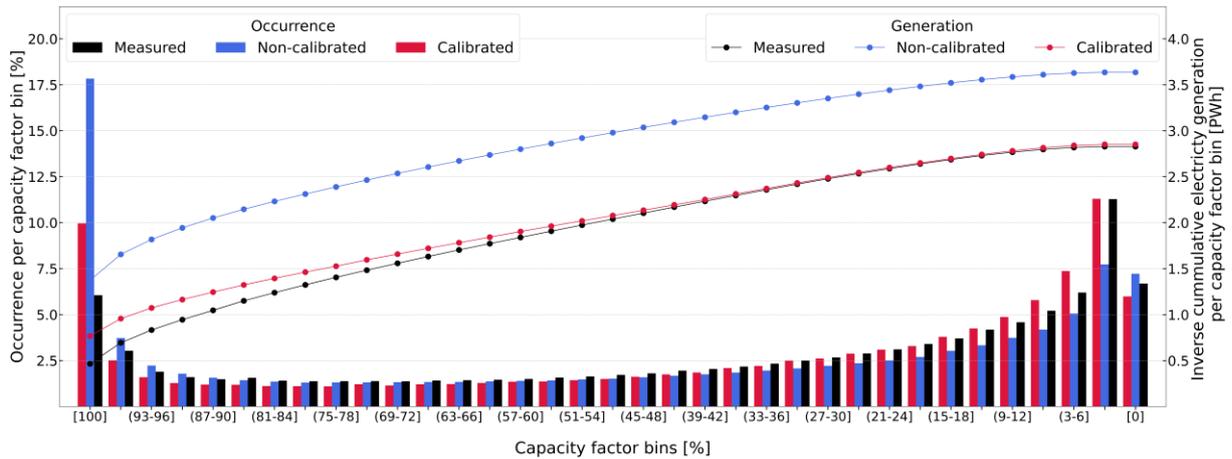

**Figure 2:** Power generation and capacity factor bin comparison between over 7.7 million hourly measurements, and calibrated and non-calibrated results. The plot illustrates capacity factor percentage bins on the x-axis, with the occurrence of capacity factors as percentages shown by bars on the left y-axis and the inverse cumulative electricity generation depicted by lines on the right y-axis.

**Spatial dimension.** The calibrated *ETHOS.RESKit$_{Wind}$* model demonstrates no significant location bias across turbine types (i.e., on- or offshore) or locations when compared to both aggregated and hourly-resolved data, reinforcing its robustness in the spatial dimension. For hourly-resolved data, most locations show a mean capacity factor error within ±10%, with a predominantly positive deviation (see Figure 3). This margin is considered acceptable for generation models. However, isolated locations in Norway and Brazil show larger mean errors (~-17%), possibly due to discrepancies between simulated turbine characteristics and measurements.

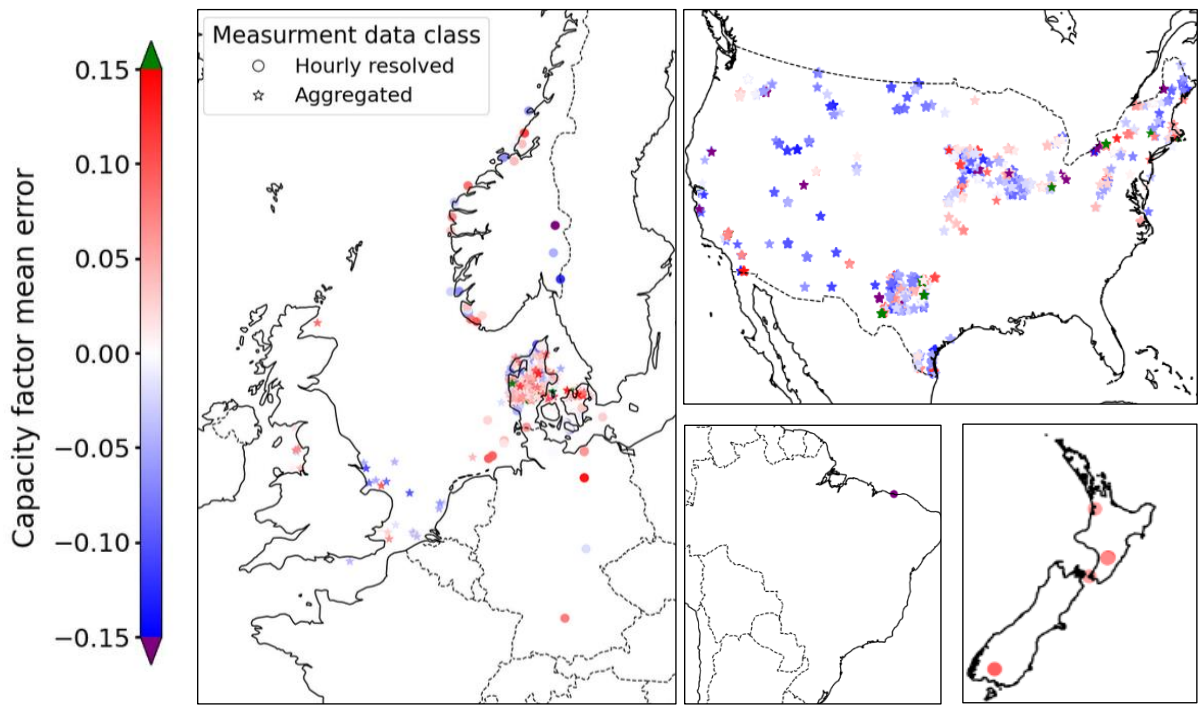

**Figure 3:** *ETHOS.RESKit$_{Wind}$* results capacity factor mean error comparison using two classes of obtained measurement data: hourly-resolved and aggregated.

Using aggregated generation data, which provides broader spatial coverage but lacks temporal detail, reveals a mix of trends (see Figure 3). Wind parks in the USA and offshore locations west of the UK show mostly negative mean errors, while those in Denmark and the UK's east coast exhibit positive deviations. While useful for expanding location coverage, this approach introduces greater uncertainty due to its lack of temporal granularity.

Most mean errors by region and turbine type fall within ±0.02, with the largest positive errors seen in New Zealand (+0.078) and Germany (+0.0526) (see Table 4 in the Supplementary material). The largest negative error occurs in Brazil (-0.175). Denmark demonstrates the most accurate results (-0.0003), followed by Norway (-0.004). No consistent discrepancies are linked to turbine types. Hourly-resolved data proves more reliable for precise analysis, enabling the identification of phenomena like induced stalling, restricted operation, and the exact onset of power generation. This enhances the model's ability to address spatial and operational dynamics effectively.

**Technological dimension.** In order to assess the efficacy of our model in replicating wind power generation, we conducted an experiment wherein we subjected the model to measured wind speeds at hub height. This enables the identification of potential input wind speed biases in temporal and location dependencies. However, reliable hub-height wind speed data is scarce. Only Denker and Wulf AG provided the requisite time-resolved wind speeds at hub height in conjunction with power generation from five distinct turbine models. Figure 4 compares the measurements and the simulation results obtained using the manufacturer's power curve included in the windpower.net [23] database and the synthetic power curve generator algorithm in *ETHOS.RESKit$_{Wind}$*.

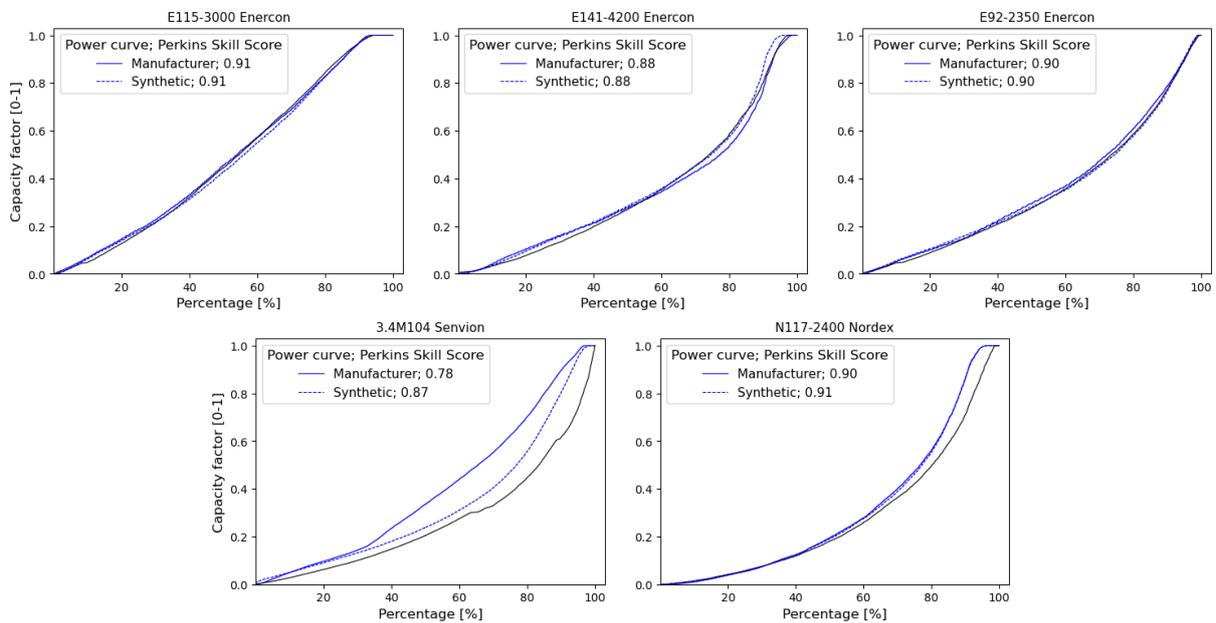

**Figure 4: Real and synthetic power curves comparison**

A comparison of the manufacturers and synthetic power curves of Enercon and N117-2400 turbines reveals a striking similarity in shape. This is corroborated by a Perkins Skill score that is highly similar in numerical terms. This indicates that the simulated power curves are highly analogous and closely aligned with the capacity factor measurements. The line plot for the 3.4M104 Senvion shows that the simulated power curves produce significantly different sorted capacity factors compared to the manufacturer's and to the actual measurements. Possible causes of the latter might come from data handling and processing of measurements or the relative developing year of the turbine (2008). A newly introduced synthetic power curve score (SPPC), see Methods section, overcomes possible data errors as well as the lack of power generation data for all turbines by comparing directy manufacturer's and syntetic power curves directy, bypassing the need to have time-resolved wind speeds at hub height. Table 3 presents the average SPCS for the turbines manufactured by the six leading producers, as reported in the Windpower.net [23] database. The data in this table demonstrate that, irrespective of the wind speed input, the synthetic power curve algorithm developed [9] and included in *ETHOS.RESKit$_{Wind}$* achieves a mean power curve score of 0.96 or higher for the majority of global installed capacity. This is especially beneficial in the case where the actual power curve is unknown.

The results obtained from all three dimensions demonstrate that the *ETHOS.RESKit$_{Wind}$* power generation model, when used in conjunction with the calibration procedure, offers a reliable assessment tool across the different measured data classes obtained. It should be noted, however, that the availability of such data on a global scale is limited, which presents a challenge to the global validation of power simulation models.

Table 3: Average synthetic power curve score in *ETHOS.RESKit$_{Wind}$* for the turbines of the top six manufacturers according to the installed capacity reported according to windpower.net [23]

| Manufacturer | Global installed capacity [GW] | Percentage of global capacity | Turbines installed [thousand] | Synthetic power curve score[1] |
|---|---|---|---|---|
| Vestas | 110.6 | 29.68 | 49.7 | 0.988 |
| Enercon | 45.4 | 12.19 | 24.2 | 0.965 |
| GE Energy | 43.0 | 11.54 | 25.0 | 0.984 |
| Siemens | 38.9 | 10.43 | 14.2 | 0.985 |
| Gamesa | 36.0 | 9.67 | 24.5 | 0.994 |
| Nordex | 21.7 | 5.81 | 8.7 | 0.992 |
| Total | 295.6 | 79.32 | 146.3 | 0.984 |

[1]the synthetic power curve score is the cumulative minimum sum of capacity factors distribution of two power curves: manufacturer and synthetic, taking as reference the manufacturer one.

## Evaluation against global wind power generation estimates

To address the limitation of global measurement data availability and evaluate the model's performance against global power estimates, *ETHOS.RESKit$_{Wind}$* was compared with publicly available power estimates at the country level for several years by the International Energy Agency (IEA) (see Figure 5). After minimizing the effects of technology differences, temporal uncertainties, and locational variations, the model showed a slight tendency to underestimate capacity factors across most countries, with discrepancies of approximately -10% or less. The IEA reports a global average capacity factor of 0.306 across 71 countries and offshore regions, while the model yielded an average of 0.278, a relative deviation of 9.1%. In comparison, the non-calibrated workflow demonstrated a significant overestimation, with an average capacity factor of 0.372 and a relative deviation of 21.3%.

Regional trends are evident. Most countries in the Americas, Oceania, East Asia, and South Africa follow the global pattern of underestimation, with exceptions like Panama, New Zealand, and Azerbaijan, which exhibit higher capacity factors. Europe presents a more varied picture. Countries around the North Sea and Sweden's offshore region show slight overestimations, which may stem from the higher density of weather masts used for wind speed calibration in these areas.

These deviations arise from several factors. The IEA dataset lacks detailed turbine technology characteristics, necessitating assumptions and external data sources to define turbine properties. Further uncertainties stem from the annual averaging of generation data, which obscures temporal dynamics, and from challenges in precisely locating turbines or identifying their commissioning dates. Additionally, external influences such as grid congestion, curtailment, import/export dynamics, and discrepancies in reporting conditions contribute to differences between simulated and actual results.

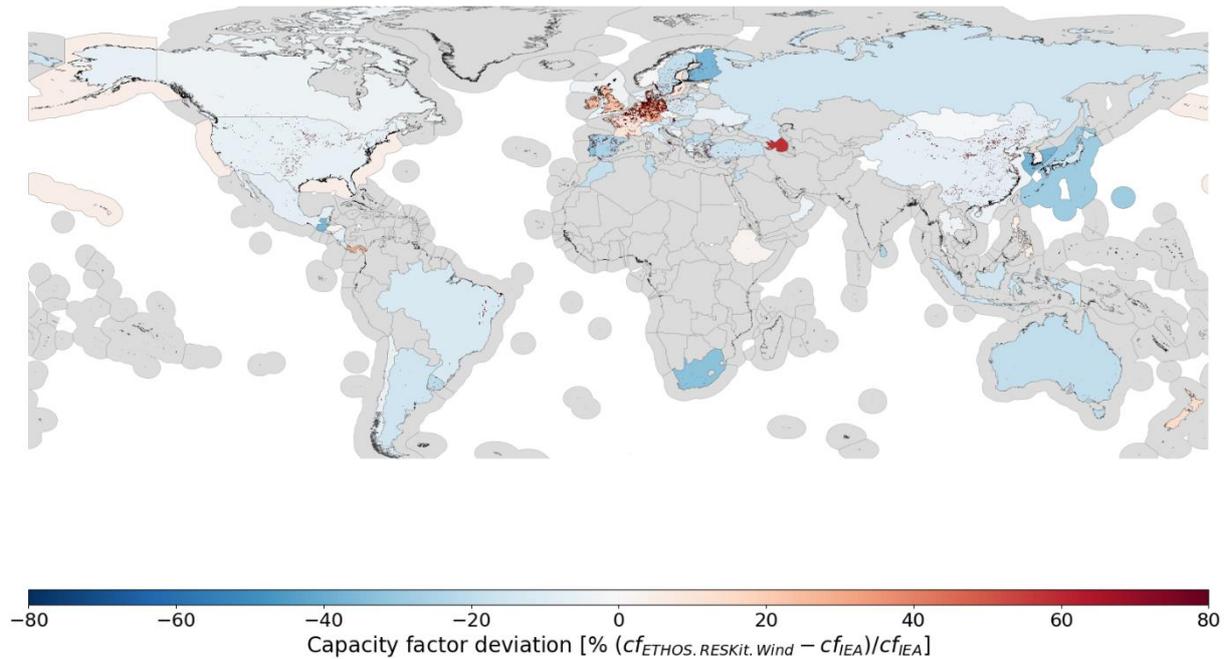

**Figure 5: Capacity factor deviation map between *ETHOS.RESKit$_{Wind}$* and IEA data [24] based on the average deviation in the years 2017 to 2021.**

The lack of globally accessible, time-resolved wind power generation data significantly hinders precision. Calibration factors, as discussed in Supplementary material 5.8, help mitigate these discrepancies, with national correction factors provided for alignment with IEA data. Furthermore, raster-format correction files extend beyond country boundaries, enabling assessments in regions without wind production and enhancing the global applicability of *ETHOS.RESKit$_{Wind}$*.

In conclusion, this evaluation underscores the model's ability to improve assessments of wind energy dependencies while highlighting the limitations of relying on aggregated country-level data. *ETHOS.RESKit$_{Wind}$* demonstrates significant advancements in accuracy compared to non-calibrated workflows, setting a strong foundation for global wind energy modeling. The previously described enhancements of our model also result in superior statistical indicators in comparison to similar models such as *renewables.ninja* (see Supplementary material 1.13).

## Discussion

In this study, we introduce *ETHOS.RESKit$_{Wind}$*, an open-source, time-resolved, validated wind power generation simulation tool designed for global applicability. *ETHOS.RESKit$_{Wind}$* leverages high-resolution wind data (250 m²) from ERA5 and GWA3, providing robust simulation capabilities and featuring the most extensive turbine model library among available tools. This library includes 880 turbine types and supports the creation of customizable synthetic power curves.

A key innovation in *ETHOS.RESKit$_{Wind}$* is its calibration process, which uses over 8 million wind speed measurements from 213 global meteorological mast sites across 25 countries and more than 8 million hours of power generation data from 152 wind turbines across seven onshore and offshore regions. This comprehensive dataset enabled a value-dependent correction of systematic wind speed biases, ensuring improved alignment with real-world data. The calibration process significantly enhances model accuracy. Temporal adjustments to input wind speeds shift capacity factors toward smaller values, aligning more closely with frequently measured capacity factors. This results in a net 0.112 improvement in capacity factor deviation compared to turbine-level time-resolved data. When simulating historical country wind fleets, the model reduced the average capacity factor deviation from 21% to 9%. Importantly, no relevant locational capacity factor deviations were observed, and the model performed consistently well across both onshore and offshore regions. Furthermore, the synthetic power curve score, which evaluates alignment between synthetic and manufacturer-provided power curves, demonstrated high accuracy. Approximately 80% of globally installed turbines achieved a minimum correlation of 0.96, underscoring the precision of the model. By reducing capacity factor deviations at both the turbine and aggregated annual levels, *ETHOS.RESKit$_{Wind}$* demonstrates superior alignment with IEA-based generation data from 71 countries. These advancements position *ETHOS.RESKit$_{Wind}$* as a leading tool for global wind energy modeling.

Importantly, although *ETHOS.RESKit$_{Wind}$* can simulate individual turbines, it is better suited to larger-scale assessments involving hundreds of turbine sites. Because of the spatial resolution characteristics of the ERA5 and GWA3 datasets, the model is less accurate at single locations where local wind speed conditions are not adequately represented. Furthermore, diurnal, seasonal, and terrain-based biases, as documented in the literature, fall outside the scope of our current correction method. Future work should concentrate on the resolution of these remaining biases in order to further enhance the precision of wind energy simulations. Moreover, enhancements in higher temporal resolution and more precise local representations of wind would be advantageous for the field. Furthermore, the entire energy and climate community would greatly benefit from the availability of more publicly accessible localized time-resolved wind speeds and power generation data. In light of these considerations, the authors urge the scientific community to engage in more collaborative endeavors and to advocate for the establishment of transparent guidelines governing the accessibility of data for scientific purposes.

The findings of this study hold substantial value for the scientific and energy system analysis communities. *ETHOS.RESKit$_{Wind}$* marks a major step forward in wind energy modeling, combining global applicability with high spatial resolution and the capability to simulate a wide range of technical turbine characteristics. As the first wind energy simulation tool to undergo a rigorous validation and calibration process across diverse spatial and temporal scales on a global level, it sets a new standard in the field. Additionally, the inclusion of regional correction factors enhances the precision of wind power assessments, even in areas currently lacking wind turbine installations. By enabling more accurate simulations, the tool equips decision-makers with critical insights to optimize renewable energy utilization and make strategic investments. This advancement significantly supports the integration of renewable energy into global power systems.

# Methods

In this section, we outline the comprehensive methodology employed for our wind energy simulation and validation approach implemented in *ETHOS.RESKit$_{Wind}$* [9,24] (see more details about *RESKit* in Supplementary material 1.6), aimed at providing robust basis for global wind energy assessments. The methodology is structured into four subsections covering (a) data acquisition and processing, (b) deriving global wind speed calibration factors aiming at addressing potential mean errors in the underlying weather data, (c) a subsequent extensive validation of our wind energy simulation workflow by comparing against time-resolved park level power generation data, country-level power generation data and national statistical data, and (d) deriving national correction factors. Each step ensures the accuracy and robustness of the employed simulation framework.

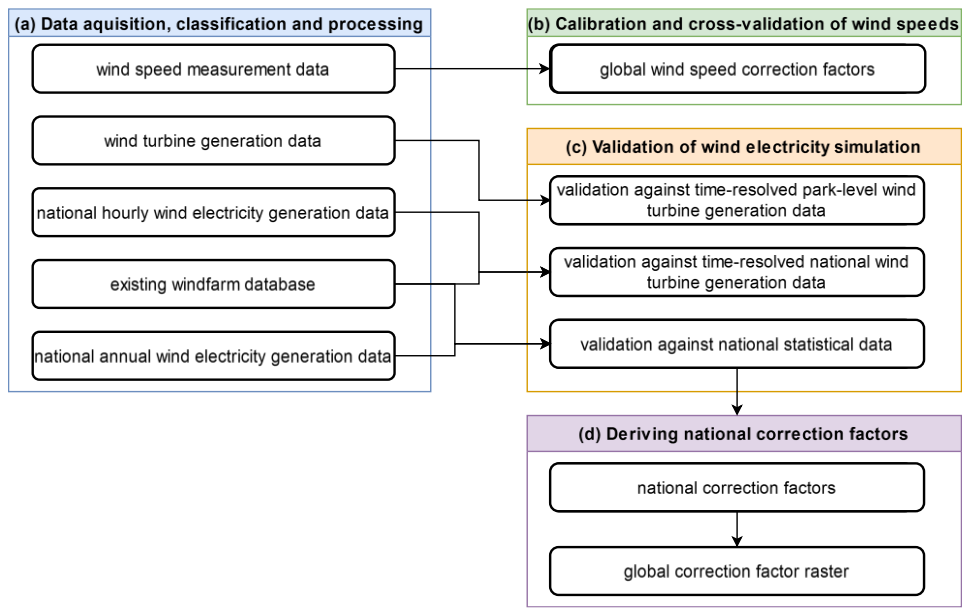

**Figure 6: Overview of the applied methodological steps.**

## Data Acquisition, Classification and Processing

In the following, we will address the acquisition, classification, and processing of data crucial for the validation and enhancement of our wind energy simulation workflow. The data sources encompass global wind speed measurements, wind turbine power generation records, information on existing windfarms, and historical national wind electricity power generation data. Each dataset plays a distinct role in refining our simulation model, either through correction or validation processes.

### Wind speed measurement data

To initiate the study, we collected 18.3 million hourly, mostly openly available recordings from 1980 to 2022 of wind speeds from meteorological masts worldwide, ranging in height from 40 m to 160 m at 210 locations in 25 countries. These recordings are utilized to derive a wind speed correction. Measurements from masts at ground level (10m) have not been included since relevant wind speed heights for turbine simulations are around 100m and large projection distances entail additional sources of error [25]. Utilizing quality control information provided together with the measured wind speeds, we filtered out erroneous measurements, e.g. no valid recording, negatives, duplicated values, etc. For further processing, we resampled the measured wind speeds and those from ERA5 to hourly values, standardized them to UTC time, and saved their geolocations as well as the measurement heights respectively.

## Wind turbine electricity power generation data

A total of 8 million hourly recordings of turbine electricity generation from 152 onshore and offshore wind turbines and wind farms from 2002 to 2021 from 6 countries globally were collected from various data sources and will be employed in a validation of our wind turbine simulation workflow. Harmonizing this data involved a process analogous to the wind speeds procedure and involved converting the power output time series to a capacity factor time series by dividing the measured power with the nominal capacity. Furthermore, in the case of wind farm data, the reported electricity output was converted into a capacity factor time series by dividing by the total park capacity.

As a quality control measure, we applied an algorithm to filter out out-of-normal operations such as curtailment, maintenance, or other irregularities from the gathered data to avoid distorting the validation results. For this, we simulated the capacity factors of the respective turbines (see Supplementary material 1.6) to first exclude observation periods in which the measured capacity factor was zero while the simulated capacity factor was greater than 0.4 to account for erroneous measurements. Second, we filtered observation periods in which the measured capacity factor exhibited zero for longer than a day to capture maintenance. Lastly, we filtered values where the measured capacity factor does not change for a minimum of 5 hours, while the difference between the measured and simulated capacity factor is greater than 0.1 to filter out curtailment lasting longer than 5 hours.

## Database of existing wind farms

Additionally, we acquired a database on existing wind farms containing data on 26,900 wind farm locations worldwide as well as databases on turbine models and power-curves [23] to simulate the existing wind fleet stock and derive national correction factors. The databases include, for instance, information on geolocation, capacity, number of turbines, hub height, turbine model, commissioning, and decommissioning dates until July 2022. It furthermore includes a turbine model database with data on the manufacturer, rated power, rotor diameter, market introduction, and minimum and maximum available hub heights of turbine models. To harmonize and check these databases, preprocessing, void-filling to estimate missing values and data filtering steps are performed as outlined in Supplementary Figure 6. Furthermore, for some entries, erroneous data was identified by manual examination. The manual examination for example involved countries with few wind farms where the capacity and capacity development of the entries in the wind farms database differed substantially from the capacity reported by the IEA Renewable Energy Progress Tracker [26]. If found to be erroneous, data on location, capacity and commissioning dates were manually corrected using additional sources such as reports, OpenStreetMap and satellite data, if possible (s. Supplementary material 1.5 for more details).

Finally, we removed locations with turbine capacities lower than 1 MW, as such turbines are comparably old and typically exhibit very low hub heights, leading to unrealistic simulation outcomes in *ETHOS.RESKit$_{Wind}$*, which is specifically designed for potential assessments of future energy systems. This arises from the substantial downscaling distance required from the 100 m ERA5 wind speed height to the turbine hub-height, introducing inherent uncertainties in the wind-speed values. In this context, such wind turbines with small hub-heights and low capacity are anticipated to have a marginal impact on the total power generation of a country due to their small capacity, justifying their exclusion from the analysis.

## Country-level statistics and time series data

We obtained annual wind power generation and capacity data from 2017 to 2021 for 71 countries and offshore regions from the IEA Renewable Energy Progress Tracker [26] as a basis for calculating national capacity factors to derive national correction factors for our simulation workflow. Data prior to 2017 has not been included as there was limited global installation of wind capacity in those years and average electricity yields are distorted by a high proportion of older, smaller turbine models. To avoid distortions in capacity factors due to capacity additions during a year, a capacity-weighted capacity factor considering monthly or even daily capacity additions was derived. This sub-annual factor was based on commissioning dates from the employed wind farm database and an extensive manual search to correct and complement the database as well as the IEA data (see Supplementary material 1.5). In Equation ( 1 ), index *i* denotes the respective wind farm, while *op_hours* is the number of hours the wind farm was operational in the respective year based on the commissioning date, and *IEA* and *WD* (wind farm database) indicate the data source.

$$cf_{country,year}^{IEA,weighted} = \frac{gen_{country,year}^{IEA}}{cap_{country,year}^{IEA}} * \frac{1}{\frac{\sum_{i,country}(ophours_{i,country}^{WD} * cap_{i,country}^{WD})}{cap_{country,year}^{WD}}}$$

( 1 )

The weighted capacity factor is especially necessary for countries with limited wind turbine capacities or a large share of commissioned capacity within a year as small deviations in the data have a large impact on the reliability of the calculated capacity factor and therefore the validation results.

In summary, Figure 7 shows the type and locations of the real-world data that were considered within this study. Statistical country values are available for various countries across the globe with data gaps predominantly in Africa, South America and South Asia. Weather mast measurements are available mainly from the USA, Europe, South-Africa and Iran while wind farm measurements are limited to the North-Sea area.

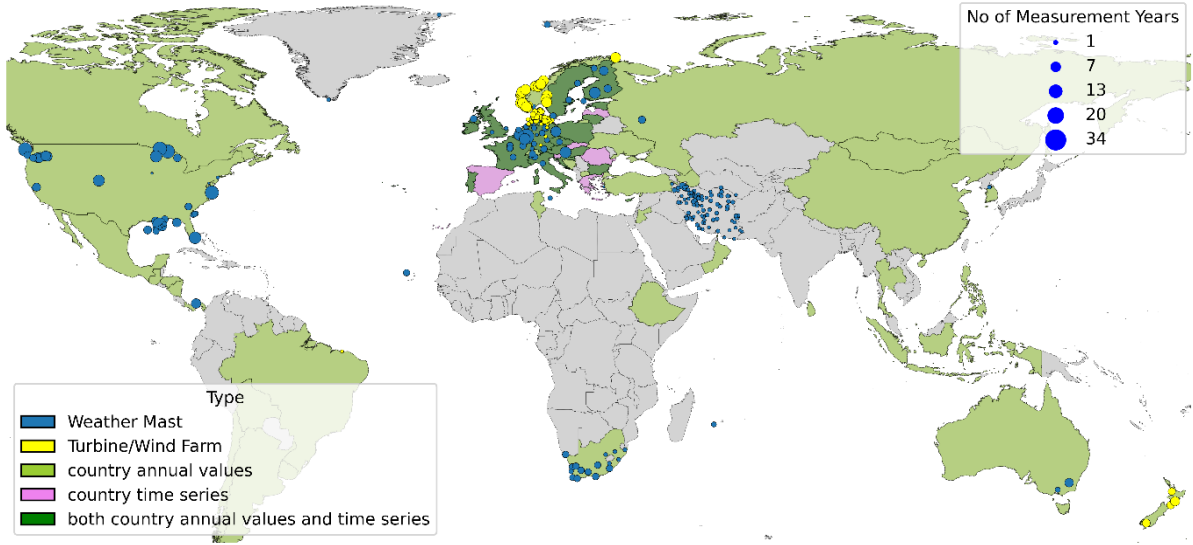

**Figure 7: Spatial overview of location and type of real-world data considered within this study. Data on specific locations, such as measurement data of weather masts (yellow) and wind turbines or wind farms (blue) are shown as circles. Data on country level such as annual country capacity factors (light green), aggregated country time series (pink) or both (dark green) are indicated by coloring the country respectively.**

## Calibration and cross-validation of estimated wind speeds from reanalysis weather data

We used the hourly measured wind speeds from meteorological masts to employ a calibration and cross-validation of the reanalysis wind speeds from ERA5 to correct for mean errors and overall under- or overestimations in the wind speed values reported by several publications [15,17,18,27]. For the calibration and cross-validation we focused on wind speeds above 2 m/s due to the operational range of wind turbines [19,28] and measurement heights between 40 and 160 m, resulting in 8.4 million hourly measurements. In a first step, we extracted the wind speeds processed within *ETHOS.RESKit$_{Wind}$* for the same locations, heights, and time periods of the weather masts without applying wake losses or any other correction factors (see Supplementary material 1.6 for a detailed description). In a second step, wind speeds were binned in 0.1 m/s categories and a proportional regression per bin was used to fit processed and measured wind speeds. Alternative regressors were tested but discarded as our tests indicated signs of overfitting or worse performance (see Supplementary material 1.7.1).

The applied proportional regression function is given in Equation ( 2 ) and is defined by a scaling factor 'a' per wind speed bin. The proportional regressor underwent fitting and validation through k-fold cross-validation. For this the data was split into 210 folds, with each fold corresponding to a mast, with the goal of assigning equal weight to each mast. The scikit-learn Python library [29] was utilized for performing the k-fold split. The choice of k-fold cross-validation is motivated by its suitability for our methodology, considering that other approaches such as the leave-one-out approach proved computationally intensive, and a rolling cross-validation performed worse than the k-fold cross-validation during initial testing.

The cross-validation procedure results in 210 fitted regressors, subsequently averaged into a single regressor from which a single scaling factor 'a' per wind speed bin is extracted. These factors were then used to correct the wind speeds within the *ETHOS.RESKit_Wind* according to Equation ( 2 ),

$$ws_{corr} = a(ws_{raw}) * ws_{raw}$$

( 2 )

where $ws_{corr}$ represents the corrected wind speed, and $ws_{raw}$ denotes the uncalibrated modeled wind speed. This unified calibration aims to rectify any general under- or overestimation present in the data. The resulting wind speed dependent scaling factors can be found in the Supplementary material. This wind speed correction is applied to every location simulated within *ETHOS.RESKit_Wind*.

To assess the quality of the regressor, we utilized a scoring function, using the mean error (ME) to account both for general deviation as well as over- or underestimation of wind speeds and capacity factors. We computed various metrics common in the literature [16,18,27,30,31] to assess the quality of the cross-validation procedure and evaluate our results. To assess the temporal correlation between measured and simulated time series, we evaluated the Pearson correlation and the detrended cross-correlation analysis (DCCA) coefficient. In addition, we used the Perkins' skill score (PSS), a probability density function, to evaluate the normal distribution. Moreover, we proposed a new probability density function called synthetic power curve score SPCS, based on the PSS from 0 to 1, where 1 represents an exact match, with the difference that it uses the cumulative minimum capacity factor distribution of two power curves, taking as reference the power curve of the manufacturer. The SPCS is described in Equation ( 3 ) where *ws* is the wind speed in each location at hub height, *Capacity factor* is the respective capacity factor distribution corresponding to wind speed *ws* for the manufacturer's and the synthetic power curve respectively.

$$SPCS = \sum_{0}^{ws} min(Capacity\ factor_{manuf,ws}, Capacity\ factor_{synth,ws})$$

( 3 )

Further analysis involves evaluating diurnal and seasonal mean errors in simulated wind speeds and reanalysis data. Results are given in Supplementary material 1.7 as the main focus of this study is the *ETHOS.RESKit_Wind* simulation workflow.

## Comparison with time-resolved wind turbine power generation data

Next, we validated the employed *ETHOS.RESKit_Wind* simulation workflow by comparing it to processed hourly measured turbine power generation data. First, *ETHOS.RESKit_Wind* was utilized to simulate the wind turbine power generation time-series for the measured time spans of each real turbine considering their specific hub heights, rotor diameters, and real power curves, if available. Where real power curves were not available, *RESKit* was used to generate turbine-specific synthetic power curves. Simulations were executed both with and without applied wind speed correction to assess the potential improvements in the simulation workflow. Furthermore, wind speed losses due to wake effects are considered using the wind efficiency curve ("*dena-mean*") from windpowerlib [32]. These wake losses were also considered for all turbine simulations in the results. Subsequently, we assessed the difference in results using various metrics, including root mean square error, DCCA coefficient, and relative mean error. These assessments occurred at the location level. Aggregated assessments are calculated by weighing each location equally when calculating metrics.

## Comparison with country-level statistical data

As the regional coverage of the available time-resolved wind turbine power generation data is limited, and our workflow is intended for global use, we first further validated and subsequently calibrated our model by comparing it against annual turbine and wind farm level power generation output and country-level annual capacity factors from the IEA [26]. The years 2017 to 2021 were used since previous years saw only limited growth of wind capacity.

Annual turbine and wind farm level power generation output from turbines in the United States, Denmark and the United Kingdom were used as they are publicly available. For each location, the average reported capacity factor was calculated using capacity and power generation. Afterwards, the locations were simulated within *ETHOS.RESKit$_{Wind}$* and the simulated capacity factor was compared with the reported capacity factor.

The filtered wind turbine database, with missing data filled in was first used to simulate country-level capacity factors by applying the method described. Second, the database was used to derive IEA-based country-level capacity factors by accounting for intra-annual capacity additions. Extensive data checks have with the following data exclusion rules have been applied: If less than 75% of the official IEA capacity is reported in the wind farm database, the corresponding year was discarded, as this indicates that the wind farm database is incomplete for that year. Omitting this year would potentially lead to large discrepancies, as a different wind fleet would be simulated compared to the one that existed in that year. Additionally, we excluded years in which the country's IEA capacity was less than or equal to 3 MW, as such a low capacity suggests a limited number of plants, where small errors in the input data could result in significant deviations in the simulated country's capacity factor. Furthermore, we discarded a country or the respective year of that country if too many entries in the wind farm database are deemed erroneous. Therefore, the number of considered years varies for each country. The list of the final countries and years considered can be found in Supplementary material 1.8. Finally, we validated the performance of our simulation workflow on a global scale by comparing the resulting, simulated annual country-level capacity factors against IEA-based country-level capacity factors by calculating the average deviation in capacity factors for every year.

Furthermore, to be able to correct our simulation workflow towards official country statistics, we derived additional correction factors, which can be optionally applied in *ETHOS.RESKit$_{Wind}$*. For this, we calculated a capacity-factor correction factor for every country, representing the average deviation in capacity factors between the IEA-based country-level capacity factors and our simulated country-level capacity factors over the years 2017 to 2021 according to Euqation ( 4 ):

$$f_{country}^{corr} = mean\left(\frac{cf_{country,year}^{IEA}}{cf_{country,year}^{RESKit}}\right).$$

( 4 )

This inverse average deviation served as a country correction factor ($f_{country}^{corr}$) implemented in *ETHOS.RESKit$_{Wind}$* to correct the electricity output. To avoid capacity factors above 1 and retain load peaks, the electricity output was corrected by adjusting the processed wind speed instead of directly correcting the simulated capacity factor. This wind speed adjusting is performed iteratively until the capacity factors match with a tolerance of 1%.

Not all countries worldwide can be covered with this approach as only a limited number of countries have installed relevant wind farm capacities. We assume that the observed deviations mostly stem from regional mean errors from which neighboring countries are also affected. Therefore, we derive a global raster of correction factors, enabling the application of global correction at any point in the world. The global raster is created by assigning every wind farm location used in this study to the respective country correction factor value and applying a global spatial interpolation over these locations. This way, existing regional mean errors are also corrected in countries without any current wind farm capacities.

## Declaration of Generative AI and AI-assisted technologies in the writing process

During the preparation of this work the authors additionally used ChatGPT, Grammarly and DeepL-Write to improve language. No content was created by AI. After using these tools, the authors reviewed and edited the content as needed and take full responsibility for the content of the publication.

## Data and code availability

The model is freely available on the institute's GitHub page (https://github.com/FZJ-IEK3-VSA/RESKit). The scripts necessary to reproduce the country-level comparisons are also included. Input data must be obtained by addressing the corresponding sources for different classes of data in question. The authors are not allowed to share this data (see Table 6).

## Acknowledgments

A major part of this work has been carried out within the framework of the H2 Atlas-Africa project (03EW0001) funded by the German Federal Ministry of Education and Research (BMBF).
Part of this work has been carried out within the framework of the HyUSPRe project which has received funding from the Fuel Cells and Hydrogen 2 Joint Undertaking (now Clean Hydrogen Partnership) under grant agreement No 101006632. This Joint Undertaking receives support from the European Union's Horizon 2020 research and innovation programme, Hydrogen Europe and Hydrogen Europe Research.
This work was partly funded by the European Union (ERC, MATERIALIZE, 101076649). Views and opinions expressed are, however, those of the authors only and do not necessarily reflect those of European Union or the European Research Council Executive Agency. Neither the European Union nor the granting authority can be held responsible for them.
This work was supported by the Helmholtz Association under the program "Energy System Design".
Open Access Publications funded by the Deutsche Forschungsgemeinschaft (DFG, German Research Foundation) – 491111487.

## Acknowledgments to data providers

We acknowledge the following people and organizations for providing dataset used in this study. We are grateful for their contribution to this work.

Henning Weisbarth - Denker & Wulf AG

## Author Contributions

Conceptualization: EUPS, PD, CW, HH; methodology: EUPS, PD, FP, CW, HH; software: EUPS, PD, CW; validation: EUPS, PD, FP, CW; formal analysis: EUPS, PD, CW; investigation: EUPS, PD, CW, HH, JW, RM, SD, SC; data curation: all named authors; writing – original draft: EUPS, PD, HH; writing – review and editing: EUPS, PD, HH, TK, JW, JL; visualization: EUPS, PD, HH; supervision: HH, JL, and DS; project administration: HH; funding acquisition: HH. All authors have read and agreed to the published version of the manuscript.

## Declaration of Competing Interest

The authors declare that they have no known competing financial interests or personal relationships that could have appeared to influence the work reported in this paper.

## References

[1] IRENA IREA. Future of wind: Deployment, investment, technology, grid integration and socio-economic aspects (A Global Energy Transformation paper). Abu Dhabi: International Renewable Energy Agency (IRENA); 2019.
[2] bp. Statistical Review of World Energy. 2022.
[3] IEA. Renewable power generation by technology in the Net Zero Scenario, 2010-2030. Renewable Power Generation by Technology in the Net Zero Scenario, 2010-2030 2022. https://www.iea.org/data-and-statistics/charts/renewable-power-generation-by-technology-in-the-net-zero-scenario-2010-2030 (accessed March 8, 2023).
[4] Schöb T, Kullmann F, Linßen J, Stolten D. The role of hydrogen for a greenhouse gas-neutral Germany by 2045. International Journal of Hydrogen Energy 2023;48:39124–37. https://doi.org/10.1016/j.ijhydene.2023.05.007.
[5] Caglayan DG, Heinrichs HU, Robinius M, Stolten D. Robust design of a future 100% renewable european energy supply system with hydrogen infrastructure. International Journal of Hydrogen Energy 2021;46:29376–90. https://doi.org/10.1016/j.ijhydene.2020.12.197.
[6] Lopion P, Markewitz P, Robinius M, Stolten D. A review of current challenges and trends in energy systems modeling. Renewable and Sustainable Energy Reviews 2018;96:156–66. https://doi.org/10.1016/j.rser.2018.07.045.
[7] McKenna R, Pfenninger S, Heinrichs H, Schmidt J, Staffell I, Bauer C, et al. High-resolution large-scale onshore wind energy assessments: A review of potential definitions, methodologies and future research needs. Renewable Energy 2022;182:659–84. https://doi.org/10.1016/j.renene.2021.10.027.
[8] Staffell I, Pfenninger S. Using bias-corrected reanalysis to simulate current and future wind power output. Energy 2016;114:1224–39. https://doi.org/10.1016/j.energy.2016.08.068.
[9] Ryberg DS, Caglayan DG, Schmitt S, Linßen J, Stolten D, Robinius M. The future of European onshore wind energy potential: Detailed distribution and simulation of advanced turbine designs. Energy 2019;182:1222–38. https://doi.org/10.1016/j.energy.2019.06.052.
[10] HoussameH, kais-siala, Buchenberg P, thushara2020, lodersky, sonercandas. tum-ens/pyGRETA: Multiple Updates 2022. https://doi.org/10.5281/zenodo.6472409.
[11] Hofmann F, Hampp J, Neumann F, Brown T, Hörsch J. atlite: A Lightweight Python Package for Calculating Renewable Power Potentials and Time Series. JOSS 2021;6:3294. https://doi.org/10.21105/joss.03294.

[12] Gelaro R, McCarty W, Suárez MJ, Todling R, Molod A, Takacs L, et al. The Modern-Era Retrospective Analysis for Research and Applications, Version 2 (MERRA-2). Journal of Climate 2017;30:5419–54. https://doi.org/10.1175/JCLI-D-16-0758.1.

[13] Davis NN, Badger J, Hahmann AN, Hansen BO, Mortensen NG, Kelly M, et al. The Global Wind Atlas: A High-Resolution Dataset of Climatologies and Associated Web-Based Application. Bulletin of the American Meteorological Society 2023;104:E1507–25. https://doi.org/10.1175/BAMS-D-21-0075.1.

[14] Hersbach H, Bell B, Berrisford P, Hirahara S, Horányi A, Muñoz-Sabater J, et al. The ERA5 global reanalysis. Quarterly Journal of the Royal Meteorological Society 2020;146:1999–2049. https://doi.org/10.1002/qj.3803.

[15] Ramon J, Lledó L, Torralba V, Soret A, Doblas-Reyes FJ. What global reanalysis best represents near-surface winds? Quarterly Journal of the Royal Meteorological Society 2019;145:3236–51. https://doi.org/10.1002/qj.3616.

[16] Gruber K, Regner P, Wehrle S, Zeyringer M, Schmidt J. Towards global validation of wind power simulations: A multi-country assessment of wind power simulation from MERRA-2 and ERA-5 reanalyses bias-corrected with the global wind atlas. Energy 2022;238:121520. https://doi.org/10.1016/j.energy.2021.121520.

[17] Pickering B, Grams CM, Pfenninger S. Sub-national variability of wind power generation in complex terrain and its correlation with large-scale meteorology. Environ Res Lett 2020;15:044025. https://doi.org/10.1088/1748-9326/ab70bd.

[18] Gualtieri G. Reliability of ERA5 Reanalysis Data for Wind Resource Assessment: A Comparison against Tall Towers. Energies 2021;14:4169. https://doi.org/10.3390/en14144169.

[19] Olauson J. ERA5: The new champion of wind power modelling? Renewable Energy 2018;126:322–31. https://doi.org/10.1016/j.renene.2018.03.056.

[20] Gualtieri G. Improving investigation of wind turbine optimal site matching through the self-organizing maps. Energy Conversion and Management 2017;143:295–311. https://doi.org/10.1016/j.enconman.2017.04.017.

[21] Davidson MR, Millstein D. Limitations of reanalysis data for wind power applications. Wind Energy 2022;25:1646–53. https://doi.org/10.1002/we.2759.

[22] Pelser T, Weinand JM, Kuckertz P, McKenna R, Linssen J, Stolten D. Reviewing accuracy & reproducibility of large-scale wind resource assessments. Advances in Applied Energy 2024;13:100158. https://doi.org/10.1016/j.adapen.2023.100158.

[23] The wind power. World wind farms database 2023.

[24] FZJ-IEK3. RESKit - Renewable Energy Simulation toolkit for Python 2023. https://github.com/FZJ-IEK3-VSA/RESKit (accessed June 9, 2023).

[25] Langreder W, Jogararu MM, Costa SA. UNCERTAINTY OF VERTICAL WIND SPEED EXTRAPOLATION n.d.

[26] Renewable Energy Progress Tracker – Data Tools. IEA 2024. https://www.iea.org/data-and-statistics/data-tools/renewables-data-explorer (accessed March 25, 2024).

[27] Brune S, Keller JD, Wahl S. Evaluation of wind speed estimates in reanalyses for wind energy applications. Adv Sci Res 2021;18:115–26. https://doi.org/10.5194/asr-18-115-2021.

[28] ENTSO-E Transparency Platform 2024. https://transparency.entsoe.eu/ (accessed March 8, 2024).

[29] Pedregosa F, Varoquaux G, Gramfort A, Michel V, Thirion B, Grisel O, et al. Scikit-learn: Machine Learning in Python. MACHINE LEARNING IN PYTHON n.d.

[30] Frank CW, Wahl S, Keller JD, Pospichal B, Hense A, Crewell S. Bias correction of a novel European reanalysis data set for solar energy applications. Solar Energy 2018;164:12–24. https://doi.org/10.1016/j.solener.2018.02.012.

[31] Perkins SE, Pitman AJ, Holbrook NJ, McAneney J. Evaluation of the AR4 Climate Models' Simulated Daily Maximum Temperature, Minimum Temperature, and Precipitation over Australia Using Probability Density Functions. Journal of Climate 2007;20:4356–76. https://doi.org/10.1175/JCLI4253.1.

[32] Haas S, Krien U, Schachler B, Bot S, kyri-petrou, Zeli V, et al. wind-python/windpowerlib: Silent Improvements 2021. https://doi.org/10.5281/zenodo.4591809.

# 1. Supplementary Material

## 1.1 Supplementary tables and figures

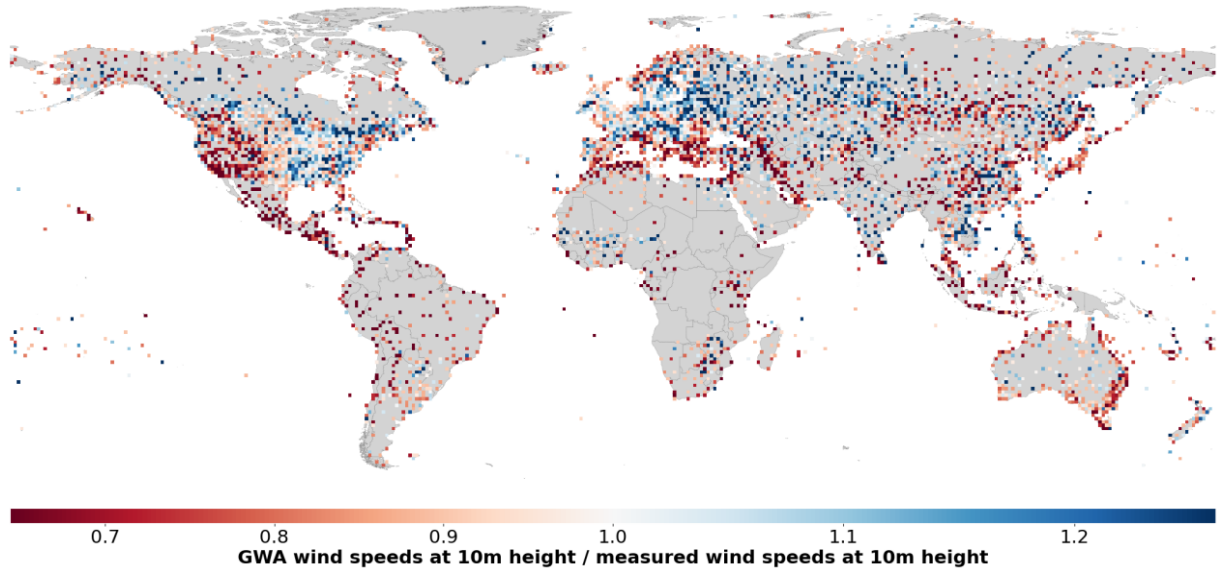

**Figure 1 Mean error observed in Global Wind Atlas and time-averaged wind speeds mean measurements at 10m height from [33,34] (own calculation)**

Table 1: Aggregated mean error by source and location

| Turbine type | Source - Location | Mean error (absolute) | Data class |
|---|---|---|---|
| Offshore | Energy Numbers - North Atlantic Ocean | -0.0128 | Aggregated |
| | Fraunhofer - Baltic Sea | 0.0139 | hourly resolved |
| | Fraunhofer - North Atlantic Ocean | 0.0487 | hourly resolved |
| Onshore | DEA - Denmark | -0.0003 | hourly resolved |
| | Denker And Wulf - Germany | 0.0526 | hourly resolved |
| | EMI – New Zealand | 0.0785 | hourly resolved |
| | NVE - Norway | -0.0042 | hourly resolved |
| | Plan- Og Landdistriktsstyrelsen - Denmark | 0.0212 | Aggregated |
| | The U.S. Wind Turbine Database - USA | -0.0223 | Aggregated |
| | UEPS - Brazil | -0.1748 | hourly resolved |

## 1.2 Wind speed measurement data

As the data formatting, level of detail and temporal resolution varies among the data sources, the meteorological data was standardized (UTC time convention, averaged one-hour resolution). Afterward, the standardized data was combined into one data set. For the meteorological data, most times quality control information provided by the data source was utilized to filter out erroneous measurements (e.g. no valid recording, negatives, duplicated values, etc.). In cases where tall masts had multiple instruments at the same height, due to the wind shadow effect, the measurement with the higher value was consistently selected.

Additionally, for the validation, wind speeds outside below 3 m/s were filtered out as they are not relevant for wind power generation due to the operational range of wind turbines.

Although the wind speed in ERA5 is stated as instantaneous, it is necessary to time-average the wind speed measurements to approximate the characteristics of the reanalysis data. While the reanalysis data provides an average over an area defined by the model's grid spacing, the measurement is a point measurement that is subject to significant local fluctuations in wind speed. Therefore, the optimal averaging period for the wind speed measurement data was investigated. The wind speed measurements averaged over 10, 20, 40, and 60-minute periods were compared to the corresponding ERA5 values using metrics such as the Pearson correlation coefficient, root mean square error (ROOT MEAN SQUARE ERROR), mean absolute error (MAE), and mean deviation shown in Figures Figure 2, Figure 3, Figure 4, Figure 5.

Nearly all metrics exhibit consistent improvement as the averaging period increases from 10 min to 60 minutes. Even higher measurement periods further improve the metrics. Nevertheless, to uphold the one-hour temporal resolution of ERA5 data, a one-hour averaging period was selected, as the benefits of longer averaging periods fail to outweigh the loss in temporal resolution.

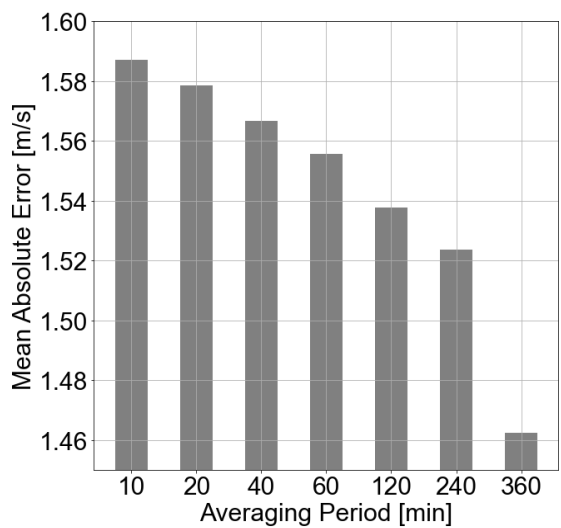

**Figure 2: Mean Absolute Error for different averaging periods.**

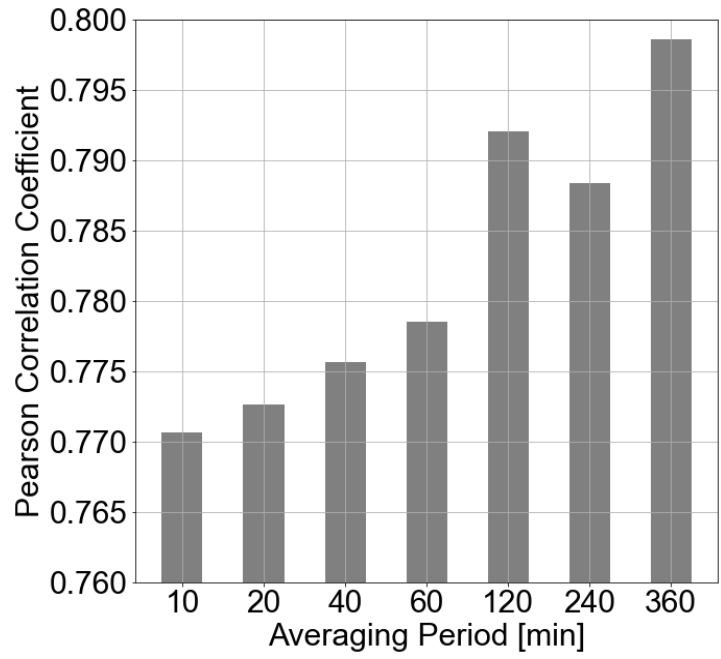

**Figure 3: Pearson correlation coefficient for different averaging periods.**

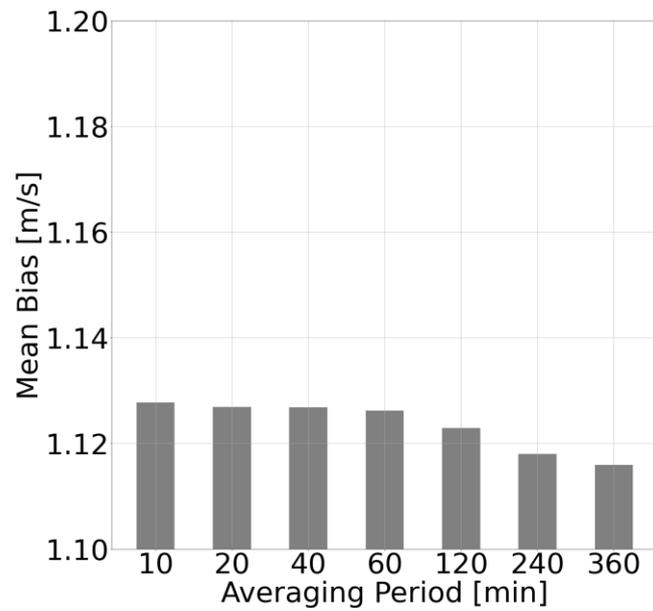

**Figure 4: Mean error for different averaging periods.**

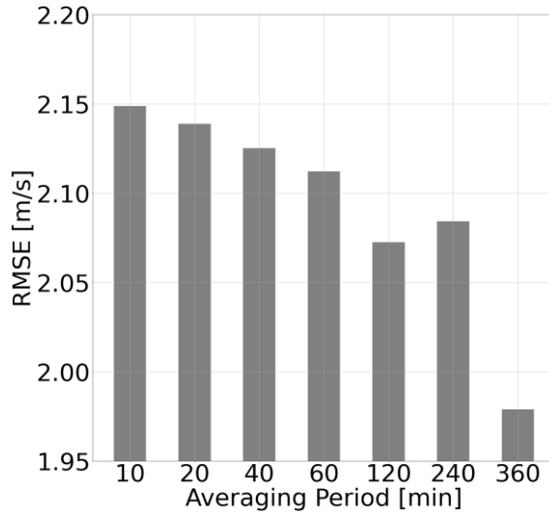

**Figure 5: RSME for different averaging periods.**

## 1.3 Wind turbine power generation data

As the data formatting, level of detail and temporal resolution varied among the data sources, the wind turbine power generation data was harmonized in a process analogous to the wind speeds procedure mentioned above. The formatted data was also combined into one data set.

Utilizing quality control information for the turbine data was not feasible since, in most cases, this information had been removed from the data. Therefore, the filtering process for shutdowns or throttling due to limitations in the power grid, maintenance, or other irregularities was more demanding. These values needed to be filtered out to avoid distorting the validation results. Therefore, the following algorithm is applied to exclude out-of-normal operations.

For this, the simulation step was already performed to be able to compare measurements and simulation. First, measurements with a measured capacity factor of zero and a simultaneously simulated capacity factor of greater than 0.4 are filtered out. Second, measurements of zero capacity factor with non-changing values for more than a day are filtered out, to capture maintenance operations. Lastly, values are filtered where the measured capacity factor does not change for a minimum of 5 hours and at the same time, the difference in simulated and measured capacity factor is greater than 0.1 to filter out curtailment operations lasting longer than 5 hours.

The processing of the turbine data requires one more step. In *RESKit.Wind*, the hourly turbine power generation is expressed as a proportion of the turbine's nominal power where the value 1 represents the operation at full capacity and 0 no power generation. The measured values are therefore converted into this form using the formula $P(NC) = \frac{P_{turbine}}{NC_{turbine}}$ with $P_{turbine}$ as the measured power, $NC_{turbine}$ as the nominal capacity and $P(NC)$ as the converted value representing the turbine load factor. Furthermore, in case the data was on wind-park level, the reported power output was averaged to a single turbine by dividing the power output by the number of turbines in the wind park.

## 1.4 Existing wind-farms database

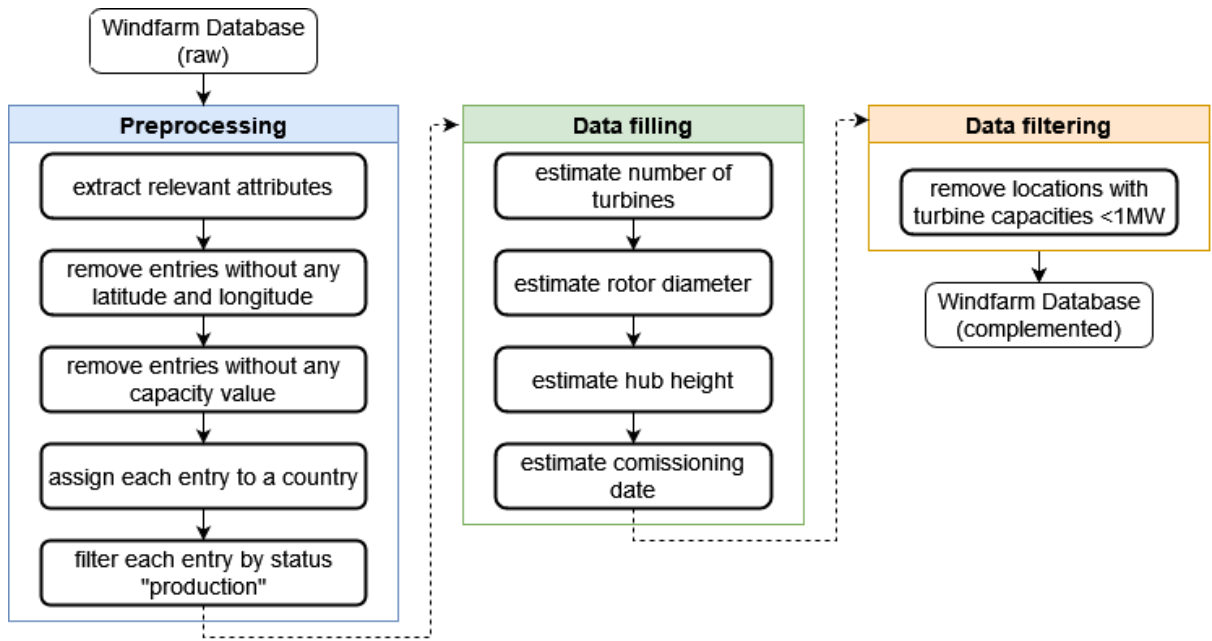

**Figure 6: Steps conducted for processing, data filling and data filtering of the wind farm database.**

Data on existing windfarms was acquired from thewindpower.net (TWP)[1] alongside databases on turbine models and power curves.

The data contains locations of 26,900 entries on operational, planned, and decommissioned onshore and offshore wind farms worldwide. The provided data attributes include the name of the wind-park, the geolocation (lat,lon), capacity, number of turbines, hub-height, decommissioning and commissioning dates as well as the used turbine model. The turbine model database contains e.g. data on the manufacturer, rated power, rotor-diameter, and market introduction as well as minimum and maximum hub height of that model.

As not all data fields were present in the wind-farm database, the following methodology was employed for filling in missing data as well as removing unusable entries as shown in Figure 6.

First, relevant data attributes were collected by matching the data of the turbine models to the wind-farms database. If the turbine model was known, the rotor-diameter was taken from the turbine model database.

Relevant attributes include "ID", "Latitude", "Longitude", "Turbine", "Manufacturer", "Hub height", "Total power", "Number of turbines", "Status", "Commissioning date", "Decommissioning date" and "Rotor diameter". Entries without any latitude and longitude data are removed. In a preprocessing step, the country and continent of each location is determined using latitude and longitude to geospatially match each location to a country shapefile. GADM[2] data was used for the administrative areas of all countries, while the exclusive economic zones (marineregions.org[3]) were used to assign offshore regions to countries. The database is filtered for wind parks in operation by filtering for status "production". Entries without a given capacity are dropped as this is the least requirement to be able to simulate the wind park.

Three parameters were estimated if they were not present for a wind farm: "Number of turbines", "Hub-height" and "Rotor diameter".

After performing the above steps, 32402 entries are present in the database. 3686 locations had missing data on the number of turbines. The number of turbines was estimated by calculating the average number of turbines per capacity based on the entries with the available number of turbines (32402-3686 entries). Multiplying this value with the capacity of the wind park yielded the number of turbines. In a second step, missing entries for the "Rotor diameter" are estimated. 7343 locations had missing entries for the rotor diameter. First, the locations were grouped by continent. Second, a power law fit was applied to fit the capacity and rotor diameter on a continent basis. Power law fit was chosen due to the observed relationship in the rest of the dataset. We refrained from using a country-level estimation as some countries (especially with few wind farms) only have a very limited amount of entries for rotor diameters. Third, missing data on the rotor diameter was estimated by applying the power law fit using the location´s capacity. In a fourth step, missing entries for the hub-height are estimated. 11549 locations had missing entries on the hub-heights. If the turbine model was given, the mean of the minimum and maximum available hub-height of the respective turbine model was used. For the remaining entries, the hub-height is estimated using a linear fit between hub-height and rotor-diameter (as the rotor-diameter showed the highest Pearson correlation of the available parameters). A linear fit was chosen because of the observed relationship on the rest of the data for locations with a hub height smaller than half the rotor diameter, the hub height is set to half the rotor diameter (To make sure the fit stays in a technical possible limit).

If the location had an unknown commissioning date and the turbine model was given, 2 years after the market introduction was assumed as the commissioning date. Finally, if the turbine model was known a power-curve from the power curve database was assigned (if available). Further, a last filtering step is applied in which locations with turbine capacities ≤1MW or park capacities ≤ 3MW are removed as such turbines typically have very low hub-heights which produce unrealistic simulation results in *RESKit.Wind*. This is primarily due to the large downscaling distance that needs to be performed from the 100m ERA5 wind speed to turbine hub-height and the resulting uncertainty in the wind-speed. However, *ETHOS.RESKit$_{Wind}$* is designed for potential assessments of future energy systems. Here, wind turbines with small hub-heights will likely not play a major role. Additionally, locations with average wind speeds ≤ 3 m/s according to GWA3 are excluded as they are considered erroneous.

## 1.5 Country-level statistical data

As outline in the methodology, annual power generation and capacity data for historical years are obtained for onshore and offshore wind on country level from the IEA. From this, a preliminary capacity factor is calculated by dividing the reported power generation for a year by the reported capacity. Notable anomalies were found in the IEA data, especially for 2022 where unrealistic capacity additions or subtractions appeared. An example of this is Indonesia in which the capacity dropped from 0.22 GW in 2020 to 0.15 GW in 2021. Therefore, 2021 was chosen as last year. Data was filtered for 2017 to 2021 as previous years only saw a limited global ramp-up of wind energy capacity. In case of significant capacity additions in a year, the preliminary capacity factor is not accurate as the newly added capacity is not generating electricity throughout the entire year. Therefore, capacity additions are weighted by the number of months the capacity addition contributed to the overall electricity generation. In case the reported country capacity from the wind park database showed significant deviations from the capacity reported by the IEA, commissioning dates were manually added by conducting internet research on individual wind farm projects using data from e.g.: power-technology.com[4]. This is especially necessary for countries with limited wind turbine capacities as small deviations in the data have a large impact on the reliability of the calculated capacity factor and therefore the validation results. Additionally, for every country we exclude years in which the country capacity in the wind-farms database is below 75% of the capacity reported by the IEA. Additionally, we exclude years in which the country's IEA capacity was below or equal 3 MW. The supplementary materials include a spreadsheet with all exclusions and corrections.

For European countries, we additionally calculated capacity-weighted time-resolved capacity factors based on country-aggregated hourly power generation values from the ENTSO-E transparency platform [5] as a further validation basis for our simulation results. As the reported capacity on the ENTSO-E transparency platform did not match the simulated country capacities in some cases, for consistency reasons the IEA capacity as ground truth for the respective year was used to calculate capacity factors. For this, the hourly power generation values were divided by the installed annual capacity reported by the IEA. It should be noted that this approach does not correctly reflect the capacity additions during a year and can therefore lead to deviations in the capacity factor. However, the data have been included as the focus of this comparison is on the correlation of the time series, which are not as prone to the aforementioned variations as the total electricity generation.

## 1.6 *RESKit$_{Wind}$* simulation workflow

The methodology in *ETHOS.RESKit$_{Wind}$* for simulating wind speed and turbine power is built upon the framework described by Ryberg et al.[6] with notable enhancements. New developments include the adoption of ERA5 data instead of MERRA-2, chosen for its superior spatial resolution and wind speed height values compared to MERRA-2. Additionally, the model incorporates the latest version of the GWA (GWA3) with an enhanced spatial resolution of 250m², a significant improvement from the 1 km² grid spacing in the original version used by Ryberg et al.[6] and Caglayan et al.[7]. We modify the long-term average used to normalize ERA5 with GWA values to be in line with the GWA3 observation period. Furthermore, the applicability of the simulation workflow is extended to global scale including offshore locations.

Part of the simulation procedure was already published in Ryberg et al.[6]. However, for comprehensiveness reasons, we present the whole workflow in Figure 7. First, all relevant turbine parameters and workflow parameters need to be specified. Turbine parameters include location, time period, hub height, rotor diameter, and capacity. Optionally, the user can provide a turbine model. If that is the case, the turbine model's power curve is used instead of a synthetic power curve[6]. It should be noted that multiple locations (up to several thousand) can be simulated at once. Optional workflow parameters include setting an availability factor, a wake reduction curve, a country correction factor, and a wind speed calibration factor.

The simulation procedure can be summarized as follows:

1. Downsampling: ERA5 wind speeds at a height of 100 m are downscaled to match the grid spacing of GWA3 using linear interpolation.
2. Long-run average (LRA): A forty-two-year average (1980-2022) is calculated based on the downscaled hourly ERA5 data at the desired location.

3. Correction factor: The LRA is divided by the value from GWA3, resulting in a correction factor. This factor is then applied to the downscaled ERA5 time series data to improve the representation of long-term orographic effects.

4. Projection: The corrected wind speeds are projected to the hub height or anemometer height using a logarithmic projection.

5. If applicable, a wind speed correction is employed

6. If applicable, wind speed losses due to wake effects are considered using the wind efficiency curves from windpowerlib[8].

7. If available, the manufacturer's power curve is used, otherwise a synthetic power curve, as described in Ryberg et al. [6], is applied.

8. To calculate the power output at the turbine location, the following steps are carried out:

    1. Air density correction: The simulated wind speed is adjusted for air density.

    2. Power curve convolution: The adjusted wind speed is convolved with the power curve using a scaling factor of 0.01 and a base factor of 0.

    3. Power output simulation: The power curve is applied to the simulated wind speeds to simulate the power output or capacity factors.

    4. If applicable, a power-output correction factor is employed (e.g. country correction factor or losses) that further corrects wind speeds to meet target capacity factor.

By default, we employ the wake reduction curve "dena_mean" from the windpowerlib python package as it showed the best alignment with our results[8]. Following Lee et al. and Fraunhofer ISI we employ an availability factor of 0.98 to approximate downtimes for e.g. maintenance of a turbine within a year[9,10]. No other losses such as environmental losses (degradation, icing etc.) are considered by default.

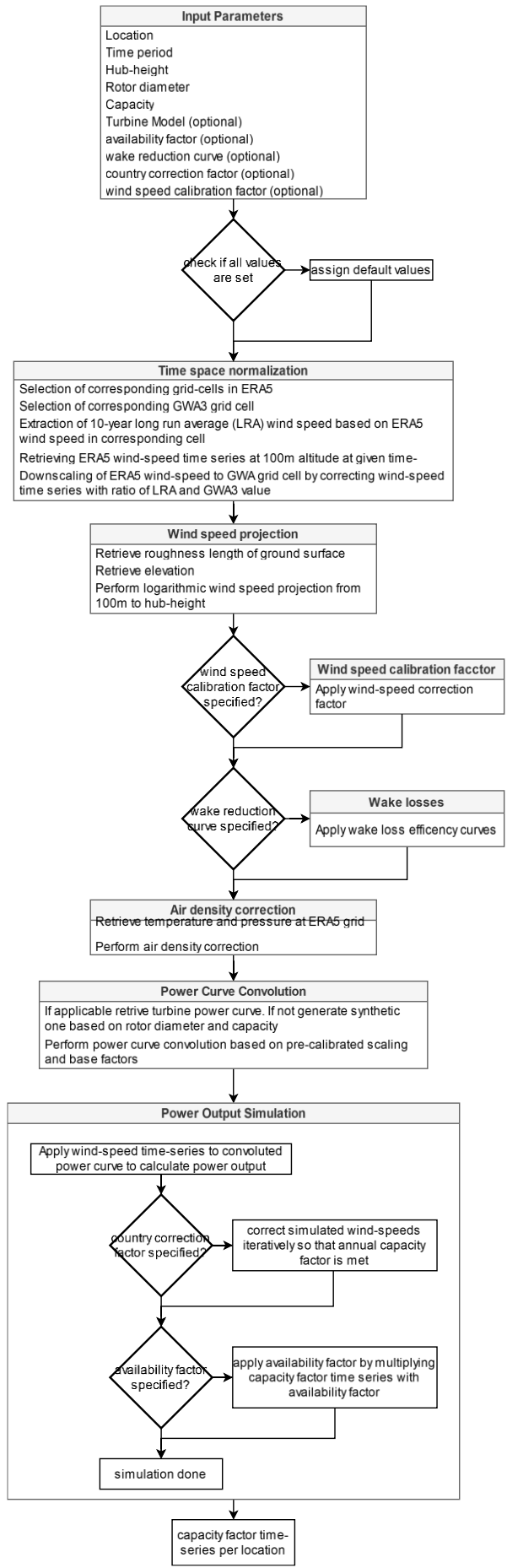

**Figure 7: ETHOS.RESKitWind power simulation workflow.**

## 1.7 Calibration and cross-validation of reanalysis wind-speed

### 1.7.1 Use of alternative cross-validation regressors

As outlined in the methodology, a k-fold cross-validation with wind speed dependent regressors is used to obtain wind-speed correction factors.

In addition to this multiple regressors were tested for wind-speed correction, such as a linear regressor, a multiple linear regressor, a multiple polynomial regressor and a Multi-Layer-Perceptron (MLP) regressor.

The linear regressor was defined by a scaling factor 'a' and offset factor 'b' and underwent the same procedure as described in the methodology. While the linear regressor showed good results on average it performed worse than the wind speed based proportional regressor. The resulting linear resulted in scaling and offset factors of *a = 0.751* and *b = 0.906 m/s*. As a result, high wind speeds were corrected strongly, leading to underestimation in high wind speed regions such as the North Sea. Therefore, the linear regressor was discarded.

The further regressors were trained on additional input data such as land cover, surface roughness, height above ground, solar elevation, month of the year as well as latitude and longitude with the goal of addressing additional spatial and temporal mean errors present in the ERA5 data.

A multiple linear regressor was tested by the same procedure outlined in the methodology. In addition to the modelled wind speed, this regressor was tested with the ESA CCI land cover code, the resulting surface roughness, height above ground, solar elevation at the given time and position, solar time, month of the year as well as latitude and longitude as additional input parameters for the regressor in all possible combinations. The best combination was found to be the ESA CCI land cover code and latitude as additional parameters for the linear correction of the modelled wind speed. The solar time is generated by the solar elevation so that it is set to hour zero with sunrise.

The second method uses a multiple polynomial regressor. Unlike the linear regressor, the polynomial regressor has the advantage of being able to correct for non-linear mean errors. While this regressor offers advanced possibilities, it also increases the risk of overfitting to the specific dataset. This regressor was fitted and validated in the same cross-validation with the addition of a multi-parameter grid search to determine the best combination of input parameters, as well as the degree of the polynomial function. The best results were obtained with the ESA-CCI landcover code, surface roughness, solar time and elevation and month of the year as parameters, and a fifth degree regressor.

The third method is the usage of a Multi-Layer-Perceptron (MLP) regressor. It was trained and validated in the same cross-validation procedure. In a grid search a wide range of hyperparameters, as well as the possible variations of input parameters was tested. As the largest deviations occur at the spatial level, latitude and longitude were initially included in the input parameters to correct for spatial mean errors.

However, it was observed that the utilization of the additional regressors did lead to notable anomalies in the corrected wind speeds showing signs of overfitting. E.g. in the case of the MLP regressor, the regressor exhibited a tendency to overly adjust the wind speed based on the specific geographic coordinates, which resulted in reduced generalization capabilities. Another issue was excessive overcorrection of particularly high wind speeds, reducing them by up to 8m/s. Hence, it was decided that the alternative regressors would not be further considered in the subsequent analysis and evaluation.

## 1.8 National correction factors

As mentioned in the Paper, the national correction factors are included in the "IEA-NationalCalibrationFactors.xlsx" data file. Moreover, the correction factor raster file are available in the ETHOS.$RESKit_{Wind}$ GitHub repository [11].

## 1.9 ENTSO-E wind power generation time-series comparison

Figure 8 compares country-level hourly power generation data from ENTSO-E and our simulations for 2017-2021 using the DCCA coefficient to assess correlation. Most countries show DCCA coefficients above 0.9, indicating a strong match with ETHOS.RESKit$_{Wind}$ results, though some coefficients drop as low as 0.69. This discrepancy primarily stems from three causes: (1) differences in wind fleet capacity between our simulations and ENTSO-E, as ENTSO-E lacks capacity time series data, leading us to rely on IEA capacity values; (2) assumptions about wind turbine characteristics, such as model type, height, and commissioning dates, which are challenging to align precisely; and (3) external factors like grid congestion, curtailment, and accounting for imports/exports introduce further differences not captured in our model. Despite these limitations, most countries display a strong alignment between simulated and ENTSO-E data, demonstrating the capability of ETHOS.RESKit$_{Wind}$ to accurately simulate country-level wind power generation These assumptions coupled with imprecise data records, challenge direct comparisons.

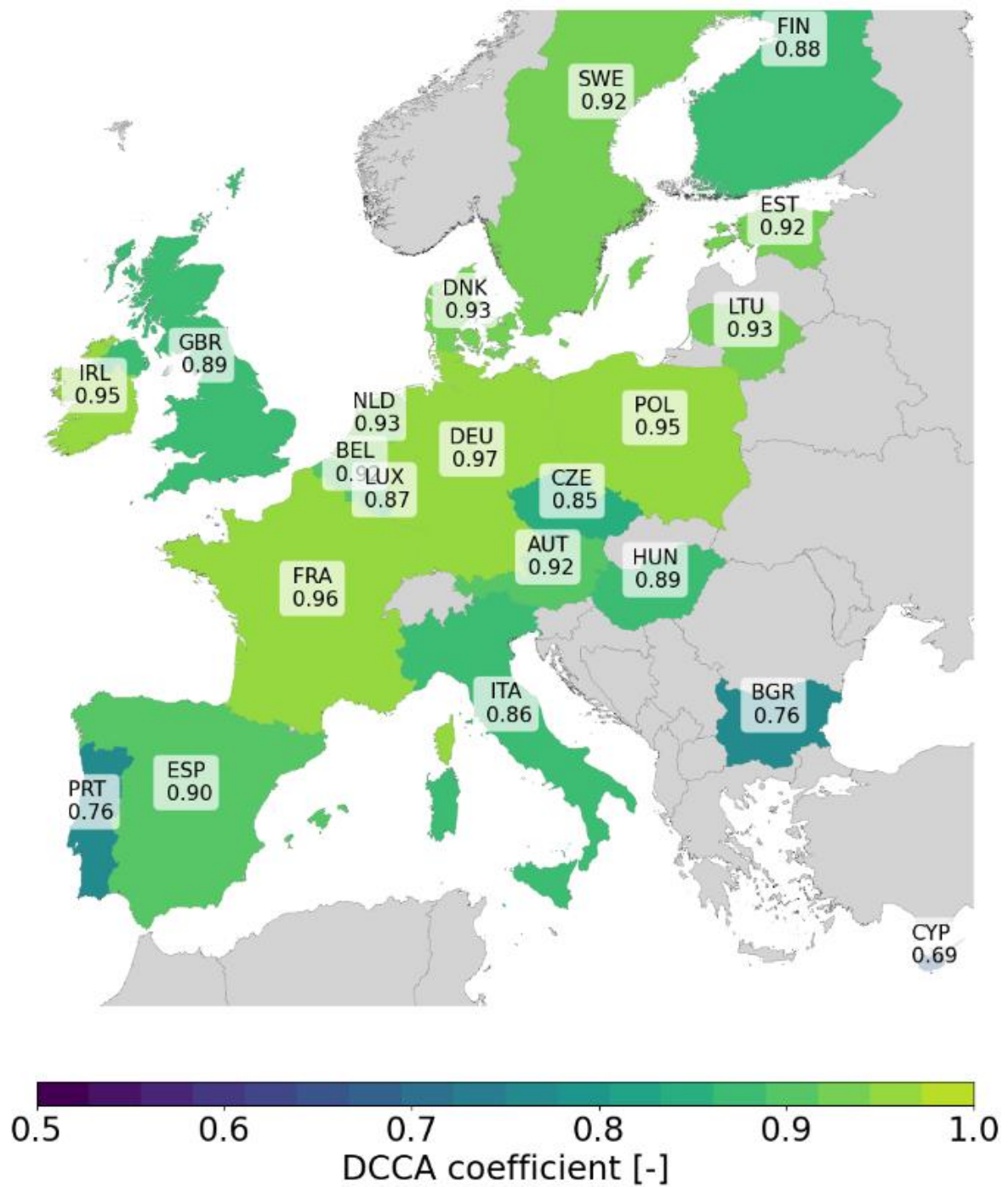

**Figure 8:** Detrended Cross-Correlation Analysis (DCCA) Coefficient between *ETHOS.RESKit$_{Wind}$* simulation and ENTSO-E publicly available data between 2017-2021.

## 1.10 Model performance limitations

In this section, we will delineate the identified limitations that were uncovered throughout our study. The intention behind this is to assist in narrowing the performance gaps that have been identified and to provide further insight into the interpretation of the results for those who are more experienced in this field. Firstly, while ETHOS.RESKitWind generally reduces the deviation in comparison to localized time-resolved wind power generation data, it is notable that there is a tendency for the generation to be underestimated in comparison to aggregated wind power generation at a country level. This discrepancy arises from the fact that aggregating a year's worth of power generation data into a single annual value for each country impedes the precise comparison of the power generation simulations. As illustrated in the results, this is particularly crucial when portraying high-capacity factors, as they account for the majority of power generation. Despite the implementation of measures to enhance the accuracy of aggregated results, it is possible that they may not fully replicate out-of-normal operations in all cases. Therefore, it is recommended that comparisons of model results be made against time-resolved measurements whenever feasible. Secondly, although wind power developers typically favor locations with favorable conditions and high wind potential, complex terrain types such as mountains, forests, and urban areas frequently result in significant discrepancies in power generation compared to these optimal locations, exhibiting both positive and negative deviations. This discrepancy may be attributed to the limitations of the Global Wind Atlas (GWA) or ERA5 data in such terrain types, as evidenced by the literature review. Additionally, diurnal and seasonal mean capacity factor deviations were observed. The impact of such wind speed mean errors is direct, affecting the performance of the model. The importance of accurately determining wind speed is evident from the wind energy power formula and is confirmed by this study. Thirdly, the limiting factor for further improvement is the availability of wind speed measurements at the hub height of the wind turbines. This is particularly relevant in the context of locational mean error correction. Should further data become available, additional regional or alternative wind speed calibration procedures may be feasible. Fourthly, it was observed that the simulation of contemporary wind turbine fleets generally produces results that are closer to the expected aggregated values for power generation. In contrast, simulations of wind turbines with a capacity below 1 MW tend to exhibit suboptimal performance, primarily due to observed discrepancies in power curve representation and the inherent difficulties in reducing wind speeds to a scale commensurate with the terrain.

## 1.11 Further recommendations for model users

The *ETHOS.RESKit$_{Wind}$* model is best suited for simulating the regional wind potential from multiple turbines distributed across the region. For more detailed simulations of individual turbines, it is recommended that other measures be employed in conjunction with the aforementioned simulations. These additional measures may include the use of weather data derived from atmospheric models at the mesoscale or microscale, local wind corrections, local environmental and operational restrictions, and so forth. In the event that the simulation results are to be compared against further measured data, it is recommended that the filtering procedures described in the Methods be undertaken in order to remove data from normal operational times. It is crucial to select the optimal wind turbine design, as this choice can significantly influence both power generation and the levelized cost of energy (LCOE). Furthermore, the wake effect or technical availability factors can be modified or omitted according to the specific simulation scenario, such as the number of turbines, the theoretical versus real performance, and so forth. In the context of country assessments, it is recommended that the results be corrected using the corresponding regional correction factors provided in supplementary material 1.8. These factors offer a comprehensive correction to account for non-physical factors influencing a region's wind power generation, based on the latest publicly available data. Depending on the specific needs of the modeler and the availability of measured data, it may be beneficial to follow a regional calibration procedure for wind speeds, based on the cross-calibration procedure presented in our study.

## 1.12 Additional discussion on the wind speed calibration procedure.

This subsection delves into additional discussion points regarding the wind speed calibration of the model, which the authors find relevant.

Firstly, it is clear throughout the results section that an accurate depiction of wind speeds is crucial for the model's performance. However, it was not possible to gather wind speeds data at wind turbine sites. As explained in section 2.1, the majority of wind speed measurements used for calibrating our workflow come from weather masts rather than actual wind turbine measurements. These masts are frequently positioned in areas characterized by intricate atmospheric conditions, such as mountainous regions, coastal areas, or urban environments, primarily to collect wind data for purposes other than assessing regions with consistent, foreseeable, or exploitable wind energy resources. Additionally, meteorological towers are not necessarily situated in areas with high wind speeds, whereas wind turbines are typically placed in regions with the highest possible wind speeds. As a result, the regressors trained on this data may not ideally match the conditions and wind speed velocities of actual wind turbine. One potential approach to better tailor the data to turbines is to exclude wind speeds from masts located in areas with low wind speeds or at measuring heights uncommon for wind turbines. However, this filtering would further reduce the amount of data used for calibration and may diminish the generalization properties of the regressor. In essence, the calibration procedure involves a tradeoff, improving performance for most locations at the expense of misrepresenting some locations. For the reasons mentioned, the potential of calibrating based on turbine data should also be investigated further. Nevertheless, the even more limited availability and quality of turbine power generation data presents the main obstacle for an approach like this. It is because of this lack of data that attention was shifted to the development of a calibration procedure focused on wind speed correction due to its potential to incorporate a larger number of measurements in diverse locations. The availability of increasingly relevant wind speeds, power generation data and turbine characteristics will enhance the calibration procedure.

Secondly, this study assessed various calibration methods, including Spline, Polynomial, and Multi-Layer Perceptron (MLP), but none yielded superior results. This lack of success may be attributed to overfitting to the wind speed data and an insufficient amount of measurement data. All tested regressors exhibited excessive reduction of wind speeds as correction at high wind speeds of 10 m/s and above, resulting in an increase in Mean Absolute Error (MAE) compared to uncalibrated values at these wind speeds. While this behavior is undesirable, it is likely due to the infrequency of these high wind speeds in the data. To address these issues, it is suggested that the correction by the regressors should be gradually reduced from this wind speed onwards. While a more sophisticated correction technique beyond linear calibration might offer greater effectiveness in mean error and error correction overall, linear calibration significantly reduced general overestimations in wind speed ranges relevant for wind turbine applications. Moreover, its quick implementation makes it suitable for large-scale usage. However, it's important to note that the linear calibration procedure should not be seen as an algorithm to "correct" ERA5 wind speed data. The authors emphasize the necessity for further advancement in reanalysis weather data models to mitigate mean errors and enhance wind energy modeling.

While investigating potential biases in ERA5 the authors utilized the HadISD dataset[12,13] (v. 3.3.0.2022f) containing wind speed measurements of global stations at 10m and compared it with ERA5 wind speeds at 10m. The DCCA coefficient was calculated for every station. When averaging the DCCA coefficients at different latitudes, a clear decrease of the DCCA coefficient towards the equator was observed as shown in Figure 9, pointing at further potential biases within ERA5 not yet addressed by literature.

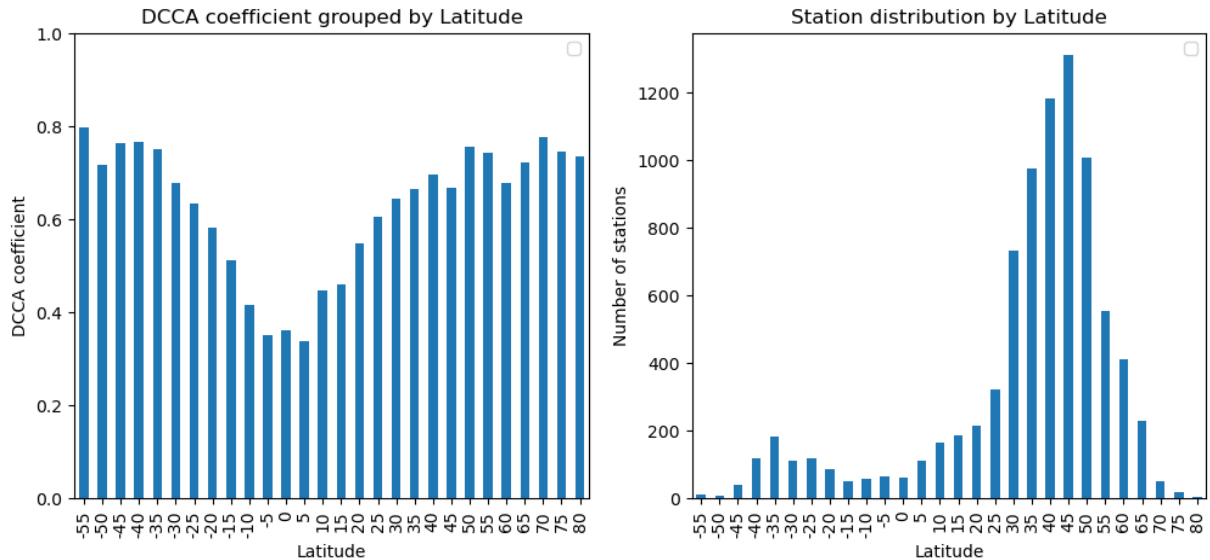

**Figure 9. DCCA coefficient and station distribution for different latitudes calculated from 10m station measurements from the HadISD dataset and 10m ERA5 wind speeds.**

Finally, the validation and calibration of model outcomes, though essential, are often secondary due to the significant time and human capital required. While the applicability of the models beyond Europe is acknowledged, their validity in other regions is not addressed. The authors recognize that overcoming this challenge can be difficult for many organizations and researchers with limited resources and time-resolved data.

## 1.13 Comparison of our model results against similar models

We evaluated our workflow *ETHOS.RESKit$_{Wind}$* against *RenewablesNinja* and *EMHIRES* [14]. It's noteworthy that while globally applicable *RenewablesNinja* operate on MERRA2, while *ETHOS.RESKit$_{Wind}$* relies on ERA5. Furthermore, we compare our results against the *EMHIRES* [14] dataset, which is limited to Europe and provides wind turbine electricity power generation time series at NUTS2 level based on MERRA2. The *atlite* model [15] and pyGRETRA were not considered in the comparison due constrains by the authors to replicate its intended performance. For the atlite model the main challenges were related to ERA5 downloading processing time and the appropriate turbine determination. *The open-source model pyGRETA* is not tailored towards simulating individual wind parks with different characteristics including turbine specific power curves. While the authors modified the source code such that single wind farms with different specifications could be simulated, it was concluded that the necessary modifications left too much room for potential errors.

Simulations of individual wind turbines and wind farms were executed across the selected tools, and subsequent results were juxtaposed against measurement data. Various metrics, such as Mean, DCCA coefficient, and Mean Error (ME), calculated per location, were employed to gauge the quality of the simulations. It is imperative to acknowledge that not all locations and measured times could be simulated with each tool by the author, only 22 locations were common amongst all the three models. Figure 10 shows that *RenewablesNinja*, *EMHIRES*, and *ETHOS.RESKit$_{Wind}$* produce mean capacity factors within 1% of the expected measurements. Figure 10 provides supplementary insight by presenting a time series evaluation. A comparative analysis of *RenewableNinjas* and *ETHOS.RESKit$_{Wind}$* demonstrate comparable value ranges, with the latter exhibiting slight improvements in both the Pearson correlation (0.02), DCCA score (0.017), Perkins skill score (0.9) and root mean square error (0.015), indicating enhanced time series correlation.

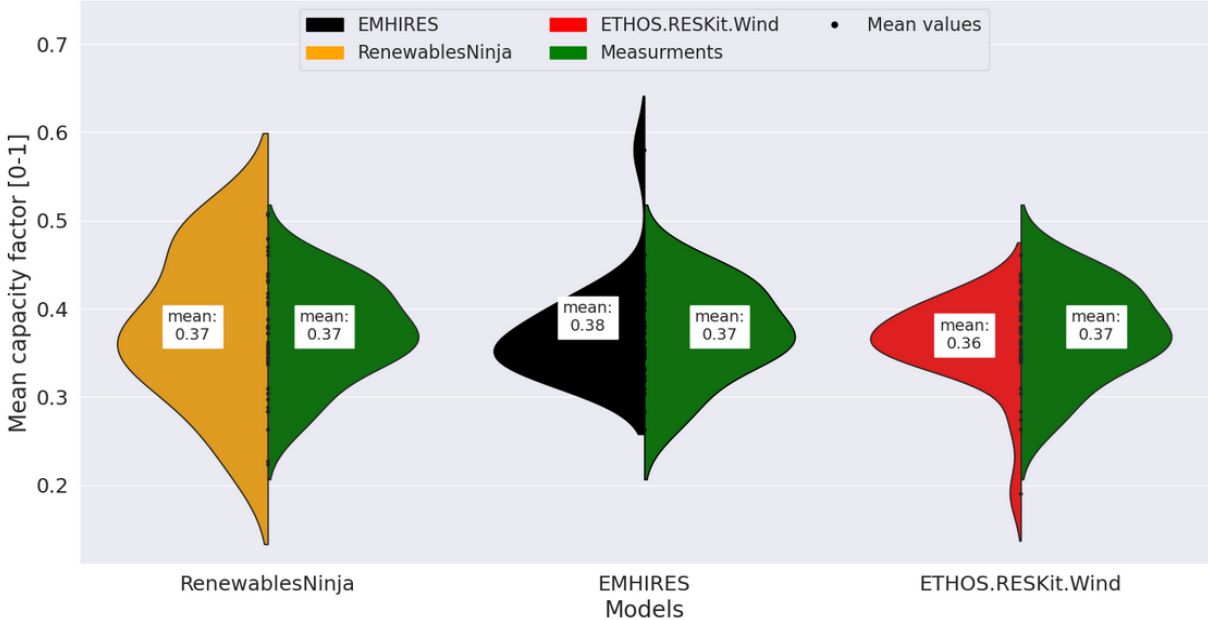

**Figure 10: Mean capacity factor comparsion between 22 locations simulated using similar models**

**Table 2: Statistical indicators comparing several hourly time series wind energy simulation model results and measurements in 21 wind parks in Europe**

| Indicator [unitless] | *EMHIRES* | *ETHOS.RESKit. Wind* | *RenewableNinjas* |
|---|---|---|---|
| **Root-mean square error** | 0.227 | 0.149 | 0.165 |
| **Pearson correlation** | 0.768 | 0.897 | 0.878 |
| **Detrended cross-correlation analysis (DCCA) coefficient** | 0.654 | 0.847 | 0.819 |
| **Perkins skill score** | 0.8663 | 0.8754 | 0.7810 |
| **Mean error** | -0.0020 | -0.0123 | 0.0031 |
| **number of locations** | 21 | | |
| **total amount observations [years]** | 98 | | |

In summary, the aforementioned findings suggest that the previously observed enhancements also result in superior statistical indicators in comparison to similar models. We declare that we have implemented each model to the best of our abilities to replicate its intended performance. However, this comparison has limitations and should not be regarded as a definitive evaluation of the models' performance.

## 1.14 Data sources

**Table 3. Data sources used in this study.**

| Data source name | Type of data | Number of locations used | period [start-end] | Original resolution | Measurement heights [min-max] | Capacity | Rotor diameter | Spatial coverage | Source |
|---|---|---|---|---|---|---|---|---|---|
| **The Tall Tower Dataset** | Meteorological Data | 174 | 1983-2021 | 10min-1h | 2-488 | n.a. | n.a. | World | Ramon et al. [16] |
| **ICOS** | Meteorological Data | 23 | 2015-today | 10min-1h | 5-341 | n.a. | n.a. | Europe | ICOS Atmosphere Level 2 data[17] |
| **NEWA** | Meteorological Data | 2 | 2016-2017 | 20Hz-10min | 60-135 | n.a. | n.a. | Europe | [18,19] |
| **Jülich Research Center** | Meteorological Data | 1 | 1981-2020 | 10 min | 100-120 | n.a. | n.a. | Germany | Personal correspondence |
| **Norwegian government agency NVE** | Turbine Data | 28 | 19 years | 1h | 46-149 | 600-5700 | 44-149 | Norway | [20] |
| **New Zealand Electricity Authority EMI** | Turbine Data | 5 | 10-16 years | 1h | 65-80 | 1600-3000 | 70-100 | New Zeeland | [21] |
| **Fraunhofer Institute** | Turbine Data | 9 | 2-3 years | 1h | 67-111 | 2300-6150 | 93-154 | Germany | Personal correspondence |
| **Danish Energy Agency** | Turbine Data | 102 | 6 years | 1h | 80-120 | 3000-8600 | 90-167 | Denmark | [22] |
| **Pedra do Sal and Beberibe Wind Farm** | Turbine Data | 20 | 2 years | 10min | 55 | 900 | 44 | Brazil | [21] ([23]) |
| **Denker and Wulf** | Turbine Data | 8 | 2 years | 1h | 128-141 | 2300-6150 | 92-141 | Germany | Personal correspondence |
| **Wind Farm Database** | Global wind farms | 26900 | n.a. | n.a. | n.a. | n.a. | n.a. | World | TWP[1] |
| **TurbineType Database** | database | n.a. | n.a. | n.a. | n.a. | n.a. | n.a. | n.a. | TWP[1] |
| **Power Curve Database** | database | n.a. | n.a. | n.a. | n.a. | n.a. | n.a. | n.a. | TWP[1] |
| **National wind power generation and capacity** | statistical data | 143 Countries | 2017-2021 | Yearly | n.a. | n.a. | n.a. | World | IEA[24] |
| **National time series of wind power generation** | time series | 23 Countries | 2017-2021 | 1h | n.a. | n.a. | n.a. | EU | ENTSOE[5] |

# References

[1] The wind power. World wind farms database 2023.
[2] GADM 2024. https://gadm.org/index.html (accessed March 25, 2024).
[3] Marine Regions 2024. https://marineregions.org/ (accessed March 25, 2024).
[4] Power Technology. Power plant profile: Jelovaca Wind Farm, Bosnia and Herzegovina. Power Technology 2022. https://www.power-technology.com/marketdata/power-plant-profile-jelovaca-wind-farm-bosnia-and-herzegovina/ (accessed March 25, 2024).
[5] ENTSO-E Transparency Platform 2024. https://transparency.entsoe.eu/ (accessed March 8, 2024).
[6] Ryberg DS, Caglayan DG, Schmitt S, Linßen J, Stolten D, Robinius M. The future of European onshore wind energy potential: Detailed distribution and simulation of advanced turbine designs. Energy 2019;182:1222–38. https://doi.org/10.1016/j.energy.2019.06.052.
[7] Caglayan DG, Ryberg DS, Heinrichs H, Linßen J, Stolten D, Robinius M. The techno-economic potential of offshore wind energy with optimized future turbine designs in Europe. Applied Energy 2019;255:113794. https://doi.org/10.1016/j.apenergy.2019.113794.
[8] Haas S, Krien U, Schachler B, Bot S, kyri-petrou, Zeli V, et al. wind-python/windpowerlib: Silent Improvements 2021. https://doi.org/10.5281/zenodo.4591809.
[9] Lee JCY, Fields MJ. An overview of wind-energy-production prediction bias, losses, and uncertainties. Wind Energ Sci 2021;6:311–65. https://doi.org/10.5194/wes-6-311-2021.
[10] Fraunhofer ISI, consentec, ifeu, TU Berlin. Langfristszenarien für die Transformation des Energiesystems in Deutschland 2022.
[11] FZJ-IEK3. RESKit - Renewable Energy Simulation toolkit for Python 2023. https://github.com/FZJ-IEK3-VSA/RESKit (accessed June 9, 2023).
[12] Dunn RJH, Willett KM, Parker DE, Mitchell L. Expanding HadISD: quality-controlled, sub-daily station data from 1931. Geoscientific Instrumentation, Methods and Data Systems 2016;5:473–91. https://doi.org/10.5194/gi-5-473-2016.
[13] Dunn RJH, Willett KM, Thorne PW, Woolley EV, Durre I, Dai A, et al. HadISD: a quality-controlled global synoptic report database for selected variables at long-term stations from 1973–2011. Climate of the Past 2012;8:1649–79. https://doi.org/10.5194/cp-8-1649-2012.
[14] European Commission. Joint Research Centre. EMHIRES dataset. Part I, Wind power generation. LU: Publications Office; 2016.
[15] Hofmann F, Hampp J, Neumann F, Brown T, Hörsch J. atlite: A Lightweight Python Package for Calculating Renewable Power Potentials and Time Series. JOSS 2021;6:3294. https://doi.org/10.21105/joss.03294.
[16] Ramon J, Lledó L, Pérez-Zanón N, Soret A, Doblas-Reyes FJ. The Tall Tower Dataset: a unique initiative to boost wind energy research. Earth System Science Data 2020;12:429–39. https://doi.org/10.5194/essd-12-429-2020.
[17] Kubistin D, Plaß-Dülmer C, Arnold S, Kneuer T, Lindauer M, Müller-Williams J, et al. ICOS Atmosphere Level 2 data, Steinkimmen, release 2023-1 2023. https://doi.org/10.18160/BJ1Z-BE0T.
[18] Dörenkämper M, Olsen BT, Witha B, Hahmann AN, Davis NN, Barcons J, et al. The Making of the New European Wind Atlas – Part 2: Production and evaluation. Geoscientific Model Development 2020;13:5079–102. https://doi.org/10.5194/gmd-13-5079-2020.
[19] Hahmann AN, Sīle T, Witha B, Davis NN, Dörenkämper M, Ezber Y, et al. The making of the New European Wind Atlas – Part 1: Model sensitivity. Geoscientific Model Development 2020;13:5053–78. https://doi.org/10.5194/gmd-13-5053-2020.
[20] Produksjonsrapporter - NVE 2024. https://www.nve.no/energi/energisystem/vindkraft/produksjonsrapporter/ (accessed March 25, 2024).
[21] Gruber K, Regner P, Wehrle S, Zeyringer M, Schmidt J. Towards global validation of wind power simulations: A multi-country assessment of wind power simulation from MERRA-2 and ERA-5 reanalyses bias-corrected with the global wind atlas. Energy 2022;238:121520. https://doi.org/10.1016/j.energy.2021.121520.
[22] Historic data of wind turbine installations in the whole of Denmark from the Danish Energy Agency (Energistyrelsen) 2019. https://doi.org/10.11583/DTU.7599698.v1.
[23] Passos J, Sakagami Y, Santos P, Haas R, Taves F. Costal operating wind farms: two datasets with concurrent SCADA, LiDAR and turbulent fluxes 2017. https://doi.org/10.5281/zenodo.1475197.

[24] Renewable Energy Progress Tracker – Data Tools. IEA 2024. https://www.iea.org/data-and-statistics/data-tools/renewables-data-explorer (accessed March 25, 2024).